\documentclass[nofootinbib,prd,twocolumn,showpacs,showkeys,preprintnumbers]{revtex4-1}
\usepackage{hyperref,amssymb,amsmath,mathrsfs,bm,graphicx}
\usepackage{multirow}
\usepackage{array}
\usepackage{dblfloatfix}
\usepackage{dblfloatfix}
\usepackage{placeins}
\usepackage{float}
\usepackage[dvipsnames]{xcolor}
\begin{document}

\title {Uncharged and charged anisotropic like--Durgapal stellar model with vanishing complexity}

\author{E. Contreras}
\email{econtreras@usfq.edu.ec}
\affiliation{Departamento de F\'isica, Colegio de Ciencias e Ingenier\'ia, Universidad San Francisco de Quito,  Quito 170901, Ecuador.\\}

\author{E. Fuenmayor }
\email{ernesto.fuenmayor@ciens.ucv.ve}
\affiliation{Centro de F\'isica Te\'orica y Computacional,\\ Escuela de F\'isica, Facultad de Ciencias, Universidad Central de Venezuela, Caracas 1050, Venezuela.\\}

\author{G. Abell\'an}
\email{gabriel.abellan@ciens.ucv.ve}
\affiliation{Centro de F\'isica Te\'orica y Computacional,\\ Escuela de F\'isica, Facultad de Ciencias, Universidad Central de Venezuela, Caracas 1050, Venezuela.\\}

\begin{abstract}
In this work we use the vanishing complexity factor as a supplementary condition to construct uncharged and charged like--Durgapal models. We provide the $g_{tt}$ component of the metric of the well-known Durgapal IV and V solutions
and a particular form for the anisotropy, related to the electric charge, to close the system of differential equations. The 
physical acceptance of the models is discussed.
\end{abstract}

\maketitle
\section{Introduction}
The use of a great variety of physical conditions in the construction of real compact configurations satisfying Einstein field equations remains a great challenge.  In some cases the material content filling the interior of the stellar object is composed by an isotropic fluid distribution where $P_r = P_{\perp}$ \cite{Delgaty:1998uy}. Although locally isotropic models constitute a very common assumption in the study of compact objects, there is strong evidence suggesting that for certain density ranges a large variety of physical phenomena of the kind we expect to find in compact objects can cause local anisotropy (see Refs. \cite{Herrera:1997plx, Herrera:2004xc, Herrera:2007kz, Glass:2013nsa, Ovalle:2017wqi, Ovalle:2019lbs, Ovalle:2017fgl, Tello-Ortiz:2020svg,Azmat:2021qig, Zubair:2020lna}, for an extensive discussion on this point). One possible source is related to the intense magnetic fields observed in compact objects such as white dwarfs, neutron stars or magnetized strange quark stars \cite{Schmidt:1995eh, Reimers:1995ia, Martinez:2003dz, PerezMartinez:2007kw, Ferrer:2010wz}. Another source is the high viscosity rates expected to be present in neutron stars, in highly dense matter produced by the opacity of matter to neutrinos in the collapse of compact objects \cite{Andersson:2004aa, Sad:2007afd, Alford:2007pj, Drago:2003wg, Jones:2001ya, vanDalen:2003uy}, and the superposition of two isotropic fluids, that allows one to evaluate the fractional anisotropy in a neutron star due to the contamination of electrons and protons required to stabilize neutrons against $\beta$ decay \cite{Bayin:1982vw}, just to name a few. It is important to remark that, although the degree of anisotropy may be small, the effects produced in compact stellar objects may be appreciable \cite{DiPrisco:1997tw}. For these reasons, the assumption of isotropic pressure is a very restrictive condition, especially in a situation where the compact object is modeled as a structure with high density (such as neutron stars, for example). Besides, the isotropic pressure condition is rendered unstable by the presence of physical factors such as dissipation, energy density inhomogeneity and shear  \cite{Herrera:2020gdg}. For these reasons, relaxing the stringent isotropic condition to allow local anisotropy within stellar matter constitutes a more realistic situation from the astrophysical point of view. In addition to the anisotropy, we may consider other possible characteristics in the internal composition of the stellar fluid as for example an electric charge. The first reported solution for charged interior configurations was given by Bonnor \cite{Bonnor:1960a} and subsequently the effects of electrical properties on compact objects have continued to be studied \cite{Florides_1983, Ray:2003gt, Ghezzi:2005iy, Boehmer:2007gq, Giuliani:2007zza, Andreasson:2008xw, MafaTakisa:2019nkj, Anninos:2001yb, Ivanov:2002jy, Barreto:2006cr,Herrera:2018czt, Sharif:2018pgq,Zubair:2021zqs,  Azmat:2021kmv}. Even though it has been said that astrophysical systems are expected to be globally charge-neutral it is expected that at certain evolutionary stages a charged astrophysical object could arise especially in transient dynamic processes. \\

From a more technical point of view, self-gravitating fluids are characterized by a number of physical variables that exceeds the number of equations provided by the theory so that additional conditions must be given in order to close the system and solve the Einstein's field equations. For instance, in the case of a static spherically symmetric anisotropic fluid, we have a set of three ordinary differential equations for the five unknown functions (metric and thermodynamic functions) so two extra conditions must be given in terms of an equation of state or an heuristic assumption involving metric and/or physical variables. In the case of considering stellar objects that have a net electric charge, obviously, we have another degree of freedom or indeterminacy that must be satisfied. In this regard, we need to provide (in addition to the boundary or initial conditions) extra information related to the local physics or restrictions on the metric variables such as the conformally flat \cite{Herrera:2001vg, herrera2014conformally} or the Karmarkar condition which allow to choose one of the metric functions as generator of the total solution \cite{karmarkar1948gravitational} (see, also \cite{ramos2021class, tello2019anisotropic, ospino2020karmarkar, maurya2017anisotropic, pant2021new, baskey2021analytical}). In the same direction, a relevant and well known concept in physics as is the complexity, applied to the scope of General Relativity, can perfectly be considered to provide the extra necessary information. 

Complexity, rather than an intuitive notion, is a physical concept that deeply relates the fundamental structure of nature and has attracted a broad spectrum of researchers in various branches of science \cite{comp3, com4, comp5, comp7, comp8, comp10}. So far, although there is not an unifying definition applied to all scenarios, several efforts have been made to provide a definition of complexity associated with the concepts of information and entropy \cite{comp11, comp12, comp13, grunwald2003kolmogorov, chapman2017complexity, yang2020time}. However, in Ref. \cite{herrera2018new} it is proposed for the first time a complexity factor for self--gravitating spheres defined by means of a structure scalar arising from the orthogonal splitting of the Riemann tensor \cite{bel1961inductions, herrera2009structure, Gomez-Lobo:2007mbg}, that manifestly exhibits that the complexity of a gravitational system is closely related to the internal structure of the object. This new approach of complexity, defined for static spherically symmetric relativistic fluid distributions , stems from the basic assumption that one of the less complex systems corresponds to a homogeneous (in the energy density) fluid distribution with isotropic pressure, assigning a zero value of the complexity factor for such a configuration. As we shall see, the vanishing complexity can be achieved also for inhomogeneous and anisotropic self--gravitating spheres so that this condition may be regarded as a non–local equation of state \cite{herrera2018new}, that can be used to obtain non-trivial configurations with zero complexity. \\


This paper is organized as follows. We dedicate the next section to introduce Einstein's equations for a spherically symmetric static and anisotropic fluid. Also some useful definitions and conventions are introduced. In section III we summarize the basics of defining the complexity factor and the vanishing complexity condition. In Sect. IV, we introduce all the variables, conventions and the equations for charged systems as well as the coupling conditions we will employ. Sects. V and VI are devoted to the study of the Durgapal IV model both uncharged and charged respectively. Sects. VII and VIII are dedicated to the study the uncharged and charged Durgapal V model. Finally, the last section is devoted to some discussions and concluding remarks.

\section{Relevant equations and conventions}\label{Einstein}

We consider a spherically symmetric distribution of static fluid, which is assumed to be locally anisotropic and bounded by a spherical surface $\Sigma$. The line element is given in Schwarzschild like coordinates by,
\begin{eqnarray} \label{metrica}
 ds^2 = e^{\nu} dt^2 - e^{\lambda} dr^2 - r^2 \left( d\theta^{2} + \sin^{2}\theta d\phi^{2}\right),
\end{eqnarray}
where $\nu$ and $\lambda$ are functions of the radial coordinate only.
The metric (\ref{metrica}) satisfy Einstein field equations given by,
\begin{eqnarray} \label{EFE}
 G^{\nu}_{\mu} = 8 \pi T^{\nu}_{\mu}.
\end{eqnarray}
The matter content of the sphere is described by the energy--momentum tensor
\begin{eqnarray}\label{energia-momentum}
T_{\mu\nu}=(\rho+P_{\perp})u_{\mu}u_{\nu}-P_{\perp}g_{\mu\nu}+(P_{r}-P_{\perp})s_{\mu}s_{\nu},
\end{eqnarray}
where, 
\begin{eqnarray}
u^{\mu}=(e^{-\nu/2},0,0,0),
\end{eqnarray}
is the four velocity of the fluid and $s^{\mu}$ is defined as
\begin{eqnarray}
s^{\mu}=(0,e^{-\lambda/2},0,0),
\end{eqnarray}
with the properties $s^{\mu}u_{\mu}=0$, $s^{\mu}s_{\mu}=-1$ (we are assuming geometric units $c=G=1$). The metric (\ref{metrica}), has to satisfy the Einstein field equations (\ref{EFE}), which are given by
\begin{eqnarray}
\rho&=&-\frac{1}{8\pi}\bigg[-\frac{1}{r^{2}}+e^{-\lambda}\left(\frac{1}{r^{2}}-\frac{\lambda'}{r}\right) \bigg],\label{ee1}\\
P_{r}&=&-\frac{1}{8\pi}\bigg[\frac{1}{r^{2}}-e^{-\lambda}\left(
\frac{1}{r^{2}}+\frac{\nu'}{r}\right)\bigg],\label{ee2}
\end{eqnarray}
\begin{equation}
P_{\perp}=\frac{1}{8\pi}\bigg[ \frac{e^{-\lambda}}{4}
\left(2\nu'' +\nu'^{2}-\lambda'\nu'+2\frac{\nu'-\lambda'}{r}
\right)\bigg]\label{ee3},
\end{equation}
where primes denote derivative with respect to $r$.

From the conservation of the energy momentum tensor it is a simple matter to find the Tolman-Oppenheimer--Volkoff
 (hydrostatic equilibrium)
 equation  for anisotropic matter, which reads
\begin{equation}
P'_r=-\frac{\nu'}{2}\left(\rho + P_r\right)+\frac{2\left(P_\bot-P_r\right)}{r}.\label{Prp}
\end{equation}
Alternatively, using 
\begin{equation}
\nu' = 2 \frac{m + 4 \pi P_r r^3}{r \left(r - 2m\right)},
\label{nuprii}
\end{equation}
which follows from the field equations, we may write (\ref{Prp}) as,
\begin{equation}
P'_r=-\frac{(m + 4 \pi P_r r^3)}{r \left(r - 2m\right)}\left(\rho+P_r\right)-\frac{2\Pi }{r},\label{ntov}
\end{equation}
where $m$ is the  mass function defined by
\begin{equation}
R^3_{232}=1-e^{-\lambda}=\frac{2m}{r},
\label{rieman}
\end{equation}
or, equivalently
\begin{equation}
m= 4\pi \int_{0}^{r} \tilde{r}^{2} \rho d\tilde{r},
\label{m2}
\end{equation}
and $\Pi$ is the local pressure anisotropy
\begin{equation}
\Pi = P_r - P_\perp.
\label{Pi}
\end{equation}

\noindent The exterior description of the spacetime is given by the Schwarzschild exterior solution \cite{schwarzschild1916gravitationsfeld}, 

\begin{equation}
ds^2= \left(1-\frac{2M}{r}\right) dt^2 - \frac{dr^2}{ \left(1-\frac{2M}{r}\right)} -
r^2 \left(d\theta^2 + \sin^2\theta d\phi^2 \right).
\label{SE}
\end{equation}

\noindent
Moreover, the matching conditions require the continuity of the first and the second fundamental form across the boundary $r=r_{\Sigma} = constant$, implying
\begin{equation}
e^{\nu_\Sigma}=1-\frac{2M}{r_\Sigma},
\label{enusigma}
\end{equation}
\begin{equation}
e^{-\lambda_\Sigma}=1-\frac{2M}{r_\Sigma},
\label{elambdasigma}
\end{equation}
\begin{equation}
[P_{r}]_{\Sigma}=0,
\label{PQ}
\end{equation}
where the subscript $\Sigma$ indicates that the quantity is
evaluated on the boundary surface $\Sigma$. Eqs. (\ref{enusigma}), (\ref{elambdasigma}) and (\ref{PQ}) are the necessary and sufficient conditions for a smooth matching of the two metrics (\ref{metrica}) and (\ref{SE}) on $\Sigma$.

\section{The Complexity Factor}\label{complexity}

This section is dedicated to summarizing the essential aspects of the definition for the complexity factor introduced in \cite{herrera2018new}. 

The Riemann tensor can be expressed through the Weyl tensor $C^{\nu}_{\alpha \beta\mu}$, the Ricci tensor $R_{\mu\nu}$ and the curvature scalar $R$, 
\begin{eqnarray} \label{Riemann1}
 R^{\nu}_{\alpha\beta\mu}&=& C^{\nu}_{\alpha \beta\mu} + \frac12 R^{\nu}_{\beta} g_{\alpha\mu} - \frac12 R_{\alpha\beta} \delta^{\nu}_{\mu}+\frac12 R_{\alpha\mu}\delta^{\nu}_{\beta}\nonumber\\&-&\frac12 R^{\nu}_{\mu} g_{\alpha\beta}-\frac16 R\left(\delta^{\nu}_{\beta}g_{\alpha\mu} - g_{\alpha\beta}\delta^{\nu}_{\mu}\right).
\end{eqnarray}
In the spherically symmetric case, the magnetic part of the Weyl tensor vanishes so we express it only in terms of its electric part:
\begin{eqnarray} \label{Weylelectrico}
 E_{\mu\nu} = C_{\mu\gamma\nu\delta}u^{\gamma}u^{\delta}.
\end{eqnarray}
Note that $E_{\mu\nu}$ may also be written as \cite{herrera2009structure}, 
\begin{eqnarray} \label{E}
E_{\mu \nu} = E \left(s_{\mu}s_{\nu} + \frac{1}{3} h_{\mu\nu}\right),
\end{eqnarray}
with
\begin{eqnarray} \label{E2}
E=-\frac{e^{-\lambda}}{4}\left[\nu '' + \frac{\nu '^{2} -\lambda ' \nu '}{2}- \frac{\nu ' - \lambda '}{r} + \frac{2(1-e^{\lambda})}{r^2}\right].\nonumber\\
\end{eqnarray}
$E_{\mu \nu}$ satisfies the following properties:
\begin{eqnarray} \label{E2}
E^{\mu}_{\;\mu} = 0 , \qquad  E_{\mu\nu} = E_{(\mu\nu)}, \qquad E_{\mu\nu}u^{\nu} = 0.
\end{eqnarray}

Now, it can be shown \cite{Gomez-Lobo:2007mbg} that the Riemann tensor may be expressed through the tensors 
\begin{eqnarray} \label{OS1}
Y_{\mu \nu}&=& R_{\mu\gamma\nu\delta} u^{\gamma} u^{\delta}\\
Z_{\mu \nu}&=& ^\ast R_{\mu\gamma\nu\delta} u^{\gamma} u^{\delta}\\
X_{\mu \nu}&=& ^\ast R^{\ast}_{\mu\gamma\nu\delta} u^{\gamma} u^{\delta}
\end{eqnarray}
in what is called the orthogonal splitting of the Riemann tensor \cite{bel1961inductions}. Here $\ast$ denotes the dual tensor, i.e. $R^{\ast}_{\mu\nu\gamma\delta}=\frac12 \eta_{\epsilon\sigma\gamma\delta} R_{\mu\nu}^{\quad\epsilon\sigma}$ and $\eta_{\mu\nu\lambda\rho}$ corresponds to the Levi--Civita tensor.

Using Einstein's field equations (\ref{EFE}) in (\ref{Riemann1}) we obtain a decomposition of the Riemann tensor given as a function of the components of the energy-momentum tensor. In \cite{herrera2018new} Eq. (\ref{energia-momentum}) was expressed in a particularly useful way so that after some manipulations (see \cite{herrera2009structure} for details), we can find explicit expressions for the tensors $Y_{\mu \nu}$, $Z_{\mu \nu}$ and $X_{\mu \nu}$ in terms of the physical variables, namely
\begin{eqnarray} \label{Y}
Y_{\mu \nu} = \frac{4\pi}{3} (\rho + 3 P) h_{\mu \nu} + 4\pi \Pi_{\mu\nu} + E_{\mu\nu},
\end{eqnarray}
\begin{eqnarray} \label{Z}
Z_{\mu \nu} = 0
\end{eqnarray}
and
\begin{eqnarray} \label{X}
X_{\mu \nu} = \frac{8\pi}{3} \rho h_{\mu\nu} +  4\pi \Pi_{\mu\nu} - E_{\mu\nu},
\end{eqnarray}
with
\begin{eqnarray} \label{OS2}
&&\Pi^{\mu}_{\nu}=\Pi \left(s^{\mu}s_{\nu} + \frac{1}{3} h^{\mu}_{\nu} \right) ; \quad P=\frac{P_{r}+ 2P_{\perp}}{3}\nonumber\\
 &&h^{\mu}_{\nu} = \delta^{\mu}_{\nu}-u^{\mu}u_{\nu}.
\end{eqnarray}
From the tensors $X_{\mu\nu}$ and $Y_{\mu\nu}$ we can define four structure scalars functions \cite{herrera2009structure} in terms of which these tensors may be written, these may be expressed as,
\begin{eqnarray} \label{XT}
X_{T} = 8\pi \rho ,
\end{eqnarray}
\begin{eqnarray} \label{XTF}
X_{TF} =  4\pi \Pi - E ,
\end{eqnarray}
\begin{eqnarray} \label{YT}
Y_{T} = 4\pi \left(\rho  + 3 P_{r} -  2 \Pi \right),
\end{eqnarray}
and
\begin{eqnarray} \label{YTF}
Y_{TF} = 4\pi \Pi + E.
\end{eqnarray}
From the above it follows that local anisotropy of pressure is determined by $X_{TF}$ and $Y_{TF}$ by
\begin{eqnarray} \label{XTF+YTF}
X_{TF} + Y_{TF} =  8 \pi \Pi ,
\end{eqnarray}
and a simple but instructive calculation performed in \cite{herrera2018new} allows us to express $Y_{TF}$ in terms of the inhomogeneity of the energy density and the local anisotropy of the system like, 
\begin{eqnarray} \label{YTF2}
Y_{TF} = 8\pi \Pi - \frac{4\pi}{r^3}\int^{r}_{0} \tilde{r}^3 \rho' d\tilde{r}.
\end{eqnarray}
Also, this last result leads us to be able to write Tolman's mass \cite{Herrera:1997plx, Herrera:1997si} as,
\begin{eqnarray} \label{m_T}
m_{T} = (m_{T})_{\Sigma}\left(\frac{r}{r_{\Sigma}}\right)^3 + r^3\int^{r_{\Sigma}}_{r} \frac{e^{( \nu + \lambda )/2}}{{\tilde{r}}} Y_{TF} d\tilde{r}.
\end{eqnarray}

Then, it is assumed that at least one of the simplest systems is represented by a homogeneous energy density distribution with isotropic pressure. For such a system the structure scalar $Y_{TF}$ vanishes. Furthermore, this single scalar function, encompasses all the modifications produced by the energy density inhomogeneity and the anisotropy of the pressure, on the active gravitational (Tolman) mass so there is a solid argument to define the complexity factor by means of this structure scalar. 

The complexity factor $Y_{TF}$, not only vanishes for the homogeneous, isotropic fluid, where the two terms in (\ref{YTF2}) vanish identically, but also for all configurations where the two terms, in the same expression, cancel each other. According to (\ref{YTF2}), the vanishing complexity factor condition, reads:
\begin{eqnarray} \label{YTF=0}
\Pi = \frac{1}{2 r^3}\int^{r}_{0} \tilde{r}^3 \rho' d\tilde{r}.
\end{eqnarray}
which may be regarded as a non–local equation of state (similar to the one proposed in \cite{Hernandez_2004}), so we can use it to impose a plausible condition on the physical variables when solving the Einstein equations. Accordingly, if we impose the condition $Y_{TF}=0$ we shall still need to provide more information in order to solve the system. 

Finally, we note that by using Einstein's equation, that (\ref{YTF2}) can be written as
\begin{eqnarray}\label{ytfcal}
Y_{TF}=\frac{e^{-\lambda } \left(\nu ' \left(r \lambda '-r \nu '+2\right)-2 r \nu ''\right)}{4 r},
\end{eqnarray}
which is summarized in terms of a relationship between the metric functions, so this condition is joined with the rest of the equations in order to integrate the system.

\section{Field equations for charged systems}

We start with the line element (\ref{metrica}) which is now associated with a static, spherically symmetric anisotropic and charged fluid distribution bounded by a surface $\Sigma$, described in Schwarzschild--like coordinates. If the fluid is charged, we must add the electromagnetic tensor $S_{\alpha\beta}$ which is given by,
\begin{eqnarray}
S_{\mu\nu} = \frac{1}{4\pi}\left(F_{\mu}^{\;\gamma}F_{\nu\gamma}- \frac{1}{4}F^{\gamma\delta}F_{\gamma\delta}g_{\mu\nu} \right)
\label{energia-momentum-em},
\end{eqnarray}
where $F_{\mu\nu}$ represents the skew–symmetric electromagnetic tensor defined as usual by,
\begin{eqnarray}
F_{\mu\nu} = \partial_{\mu}A_{\nu} - \partial_{\nu}A_{\mu},
\label{Fmunu}
\end{eqnarray}
satisfying Maxwell equations,
\begin{eqnarray}
F^{\mu\nu}_{\quad ;\nu} = 4 \pi J^{\mu},
\label{maxwell}
\end{eqnarray}
where $A_{\mu}$ is the four potential and $J_{\mu}$ the four current. In the static and spherically symmetric case, they are expressed as, 
\begin{eqnarray}
A^{\mu}=\Phi (r) \delta^{\mu}_0 \quad,\quad J^{\mu}= \sigma (r) u^{\mu} 
\label{Amunu}
\end{eqnarray}
where $\Phi$ is the electric scalar potential and $\sigma$ the charge density (both functions of the coordinate $r$). The electric charge interior to radius $r$ has been defined by using the relativistic Gauss’s law as follows \cite{Bekenstein:1971ej},
\begin{eqnarray}
q(r) = 4 \pi \int_{0}^{r} \sigma(\tilde{r}) \tilde{r}^{2} e^{\lambda/2} d\tilde{r},
\label{cargainterna}
\end{eqnarray}
implying charge conservation. As we are dealing with a spherically symmetric, static configuration, this implies that the only non–vanishing components of the electromagnetic tensor $F^{\mu\nu}$ are $F^{01}$, which correspond precisely to the electric field $E(r)$ along the radial direction, namely
\begin{eqnarray}
E=  \frac{q(r)}{r^2} e^{- (\lambda + \nu)/2}.
\label{campoE}
\end{eqnarray}

From now on we consider that the energy distribution within the fluid corresponds to an anisotropic fluid, given in (\ref{energia-momentum}), coupled to an electromagnetic field (\ref{energia-momentum-em}). The metric must satisfy Einstein--Maxwell equations for the interior spacetime,
\begin{eqnarray}
G^{\mu}_{\nu} = 8 \pi (T^{\mu}_{\nu} + S^{\mu}_{\nu} ),
\label{E-M}
\end{eqnarray}
that in this case may be written as 
\begin{eqnarray}
\kappa\rho+\frac{q^{2}}{r^{4}}&=&\frac{\lambda'}{r}e^{-\lambda}+\frac{1}{r^{2}}(1-e^{-\lambda}),\label{e1}\\
\kappa P_{r}-\frac{q^{2}}{r^{4}}&=&
\frac{\nu'}{r}e^{-\lambda}-\frac{1}{r^{2}}(1-e^{-\lambda}),\label{e2}\\
\kappa P_{\perp}+\frac{q^{2}}{r^{4}}&=&
e^{-\lambda}\left(\frac{\nu''}{2}
-\frac{\lambda'\nu'}{4}+\frac{\nu'^{2}}{4}-\frac{\nu'-\lambda'}{2r}
\right).\label{e3}
\end{eqnarray}
It is clear that the uncharged solution is recovered when $q=0$. Next, the Misner and Sharp mass function $m(r)$ (\ref{rieman}) can be generalized to include the electromagnetic contribution by \cite{Herrera:2011cr},
\begin{eqnarray}
m = \frac{r^3}{2} R_{23}^{\quad 23} + \frac{q^2}{2r}= 4\pi \int_{0}^{r} \tilde{r}^{2} T^0_0 d\tilde{r},
\label{m3}
\end{eqnarray}
where $T^0_0$ is given by the left side of (\ref{e1}) in this case. So when considering the total mass of the compact structure, if the fluid distribution contains electric charge, it will increases by a certain amount, and this effect is provided by the electric field. This implies that the surface gravitational red–shift is also altered and one could try to check the models using some observational data.

 Note that Eqs. (\ref{e1}), (\ref{e2}) and (\ref{e3}) corresponds to a system of three independent equations with six unknowns, namely $\{\nu,\lambda,\rho,P_{r},P_{\perp},q\}$ so we need to provide three conditions in order to solve the system. In this work we shall provide the $g_{tt}$ metric component of the well known Durgapal IV and Durgapal V solutions, the vanishing complexity condition and a particular profile for the anisotropy of non--vanishing charge solutions.

In order to find a complete and consistent stellar model, one needs to satisfy the matching conditions in a smoothly way at the surface interface $\Sigma$ (that represents the compact object surface) with the external space–time which we assume as the Reissner--Nordstr\"om spacetime,
\begin{eqnarray}\label{SchExt}
   ds^2 &=& \left( 1 - \frac{2M}{r} + \frac{Q^2}{r^2} \right) dt^2 - \left( 1 - \frac{2M}{r} + \frac{Q^2}{r^2} \right)^{-1}\!\!dr^2 \nonumber\\
    & & - r^2 (d\theta^2 + \sin^2\theta d\phi^2)\, ,
\end{eqnarray}
where $M$ and $Q$ denote the total mass and charge respectively. Moreover, the matching conditions require \cite{Israel:1966rt}
\begin{eqnarray}
    e^{\nu_\Sigma} &=& 1 - \frac{2M}{r_\Sigma} + \frac{Q^2}{r_\Sigma^2}\,, \label{MC01} \\
    e^{-\lambda_\Sigma} &=& 1 - \frac{2M}{r_\Sigma} + \frac{Q^2}{r_\Sigma^2}\,, \label{MC02}\\
    P_r(r_\Sigma) &=& 0\,, \label{MC03}\\
    q(r_\Sigma) &=& Q\,. \label{MC04}
\end{eqnarray}
A null radial pressure at the boundary surface $\Sigma$ is a necessary mechanism to confine the matter content inside a bound space– time region, which in turn determines the size of the configuration. 

It is opportune to clarify that the same expression presented for the complexity factor $Y_{TF}$ and the non-local state equation that defines the vanishing complexity of the system, for the non-trivial case (\ref{YTF=0}), is perfectly valid in the general case of an electrically charged fluid distribution. The complexity factor for a charged (spherically symmetric) fluid is known, because the structure constants in that case have also been found (see \cite{Herrera:2011cr} for details and discussion). The inclusion of electric charge results in defining ``effective'' thermodynamic matter variables that ``absorb'' the contribution of the electric charge. The effective variables are just the corresponding ordinary variables with all contributions (electric charge) included.


\section{Durgapal IV solution using vanishing complexity}

Let us first consider the uncharged case. For this we set $q=0$ in Eqs. \eqref{e1}, \eqref{e2} and \eqref{e3}. We use the $g_{tt}$ metric coefficient as a seed known solution
\begin{eqnarray}\label{nu}
e^{\nu}=A \left(1+c r^2\right)^4\, ,
\end{eqnarray}
where $A$ and $c$ are constants. The vanishing complexity condition \eqref{YTF=0} enables us to find the radial metric coefficient
\begin{eqnarray}\label{lambda}
e^{\lambda}=(1+c r^{2})^{2}\,.
\end{eqnarray}
Replacing both metric coefficients in Eqs. \eqref{e1}, \eqref{e2} and \eqref{e3} (with $q=0$) we find
\begin{eqnarray}
\rho &=& \frac{c \left[6 + c r^2 \left(c r^2+3\right)\right]}{8 \pi  \left(1 + c r^2\right)^3},\\
P_{r} &=& \frac{c \left[6 - c r^2 \left(c r^2+3\right)\right]}{8 \pi  \left(1 + c r^2\right)^3}, \label{pradial}\\
P_{\perp} &=& \frac{6 c}{8 \pi  \left(1 + c r^2\right)^3}\label{ptangencial}\,.
\end{eqnarray}
Using equations (\ref{pradial}) and (\ref{ptangencial}) or just evaluating \eqref{YTF=0} we obtain directly
\begin{equation}\label{aniso}
    \Delta = \frac{c^2 r^2(3+cr^2)}{8\pi (1+cr^2)^3}\,.
\end{equation}
Now, we proceed with the matching conditions \eqref{MC01}, \eqref{MC02} and \eqref{MC03} (with $Q=0$) and find that
\begin{eqnarray}
A &=& \frac{64}{(1 - \sqrt{33})^6}\;,\\
M &=& \frac{15-\sqrt{33}}{(1-\sqrt{33})^2}R \;\label{compacidad1},\\
c &=& \frac{-3 + \sqrt{33}}{2R^2}\;.
\end{eqnarray}
It is worth noticing that from Eq. (\ref{compacidad1}) the compactness parameter ($M/R$) is 0.4112 which corresponds to a very compact object. Besides, the matching conditions exhaust the free parameters of the model so we do not have any parameter that modulates the anisotropy of the system (in other words if $R$ is given then also $c$ is given).\\

\begin{figure}[h!]
\centering
\includegraphics[scale=0.5]{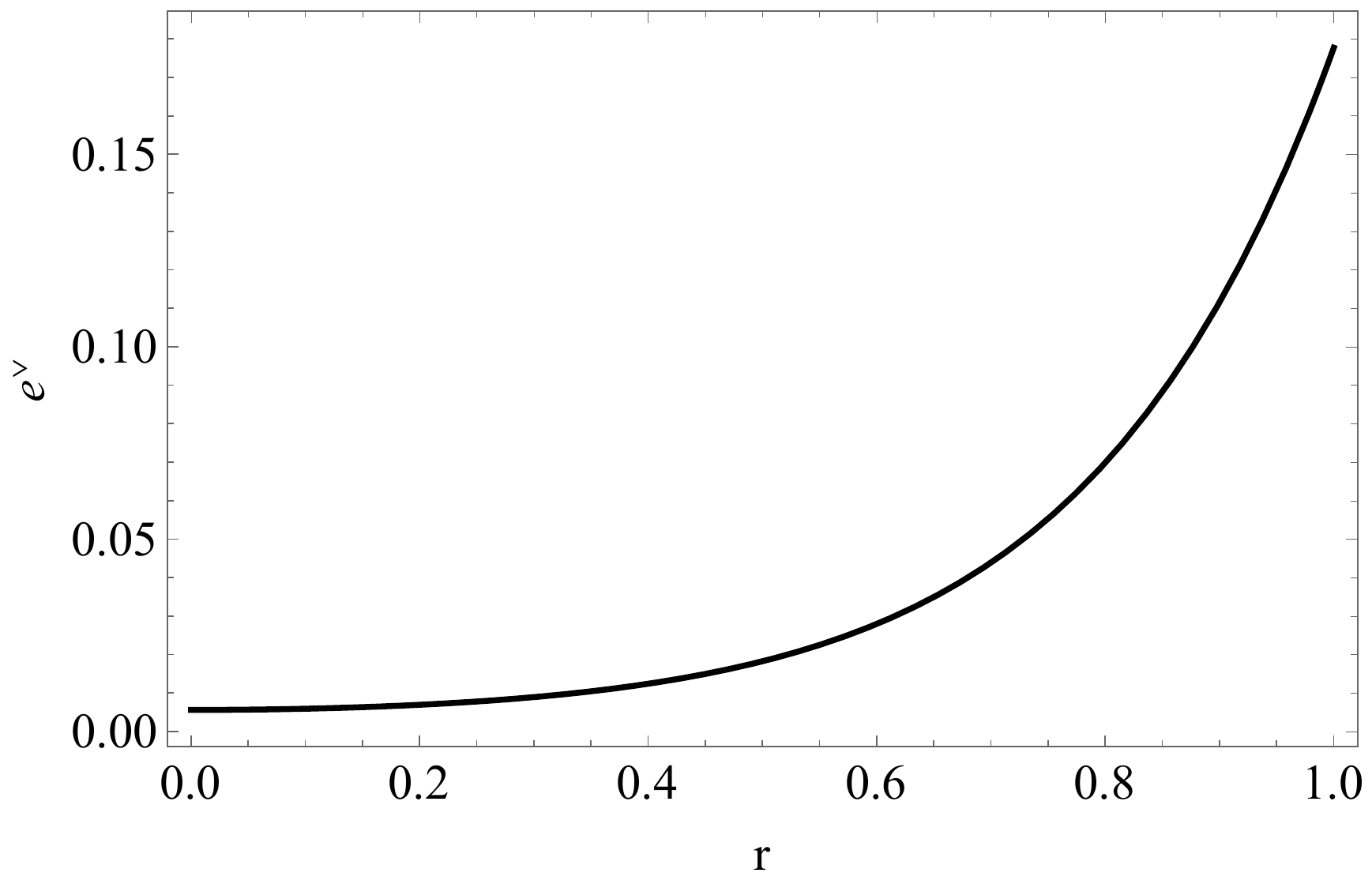}
\caption{\label{fenu1}
Metric $e^{\nu}$ as a function of the radial coordinate $r$ with $R=1$ for Durgapal IV ($q=0$) solution. 
}
\end{figure}

\begin{figure}[h!]
\centering
\includegraphics[scale=0.5]{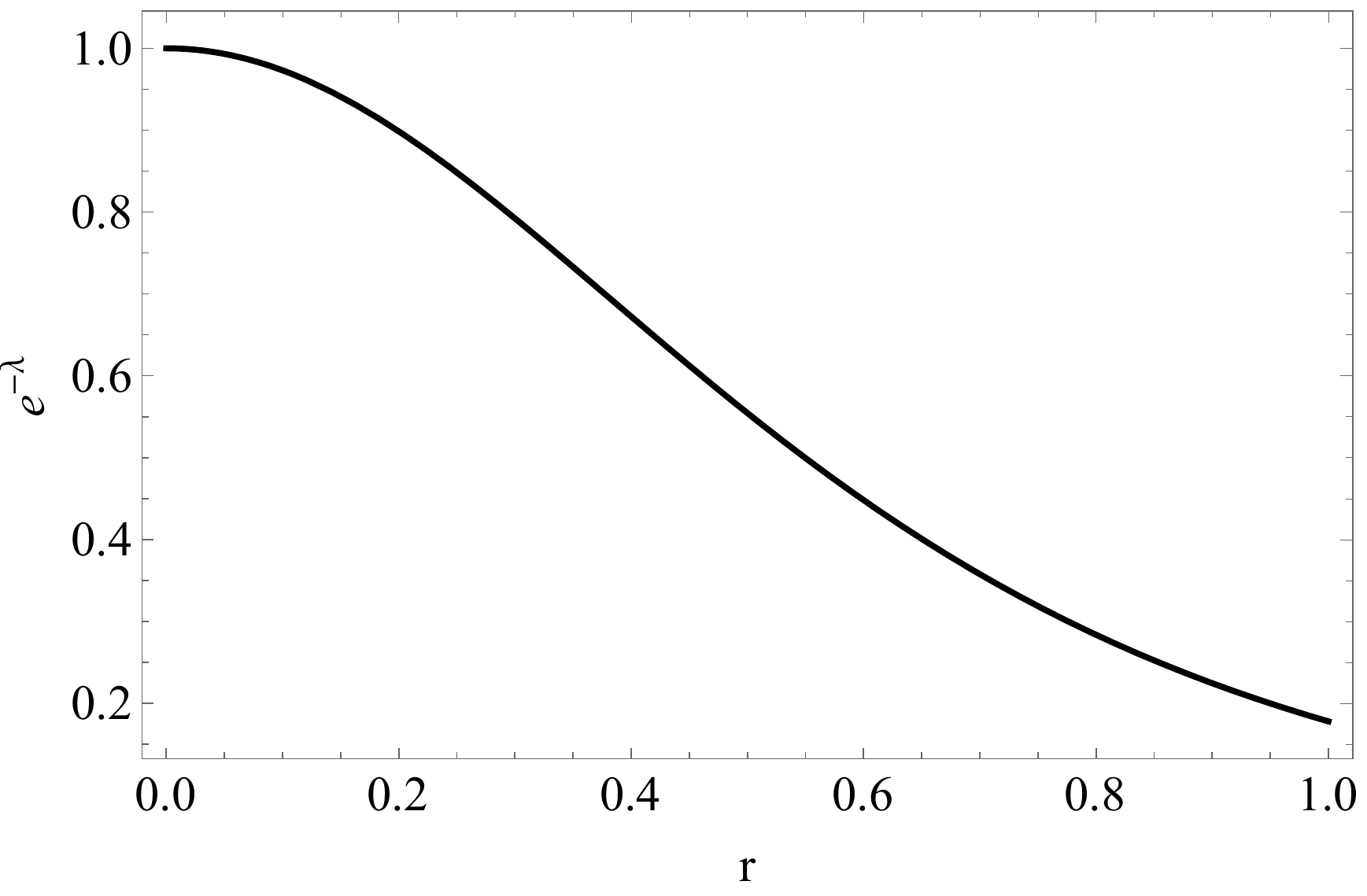}
\caption{\label{felambda1}
Metric $e^{-\lambda}$ as a function of the radial coordinate $r$ with $R=1$ for Durgapal IV ($q=0$) solution. 
}
\end{figure}

In in figures (\ref{fenu1}) and (\ref{felambda1}) we show the metric functions $e^\nu$ and $e^{-\lambda}$ as a function of the radial coordinate. Note that both are positive, finite and free of singularities, as expected. Moreover, at the centre, they satisfy $e^{-\lambda(0)}=1$ and $e^{\nu(0)}=const$. From this we can observe a monotonously increasing correct behavior for the mass function, this is consistent with the adequate behavior of the metric function $\lambda$ 
\cite{Hernandez:2020pcn, Suarez-Urango:2021cjy,Suarez-Urango:2021mjg}. 

\begin{figure}[h!]
\centering
\includegraphics[scale=0.5]{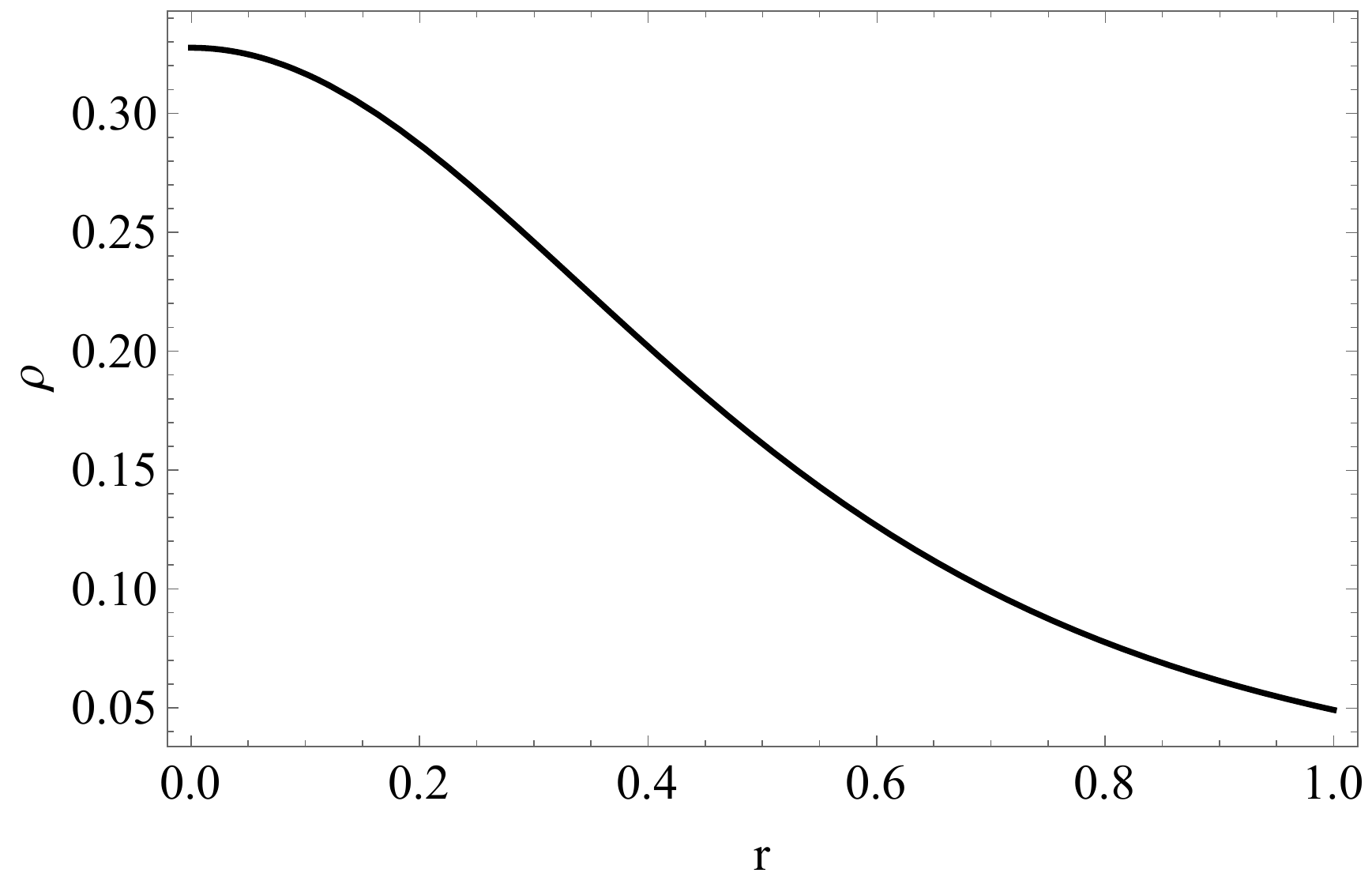}
\caption{\label{fdensidad1}
$\rho$ as a function of the radial coordinate $r$ with $R=1$ for Durgapal IV ($q=0$) solution. 
}
\end{figure}

\begin{figure}[h!]
\centering
\includegraphics[scale=0.5]{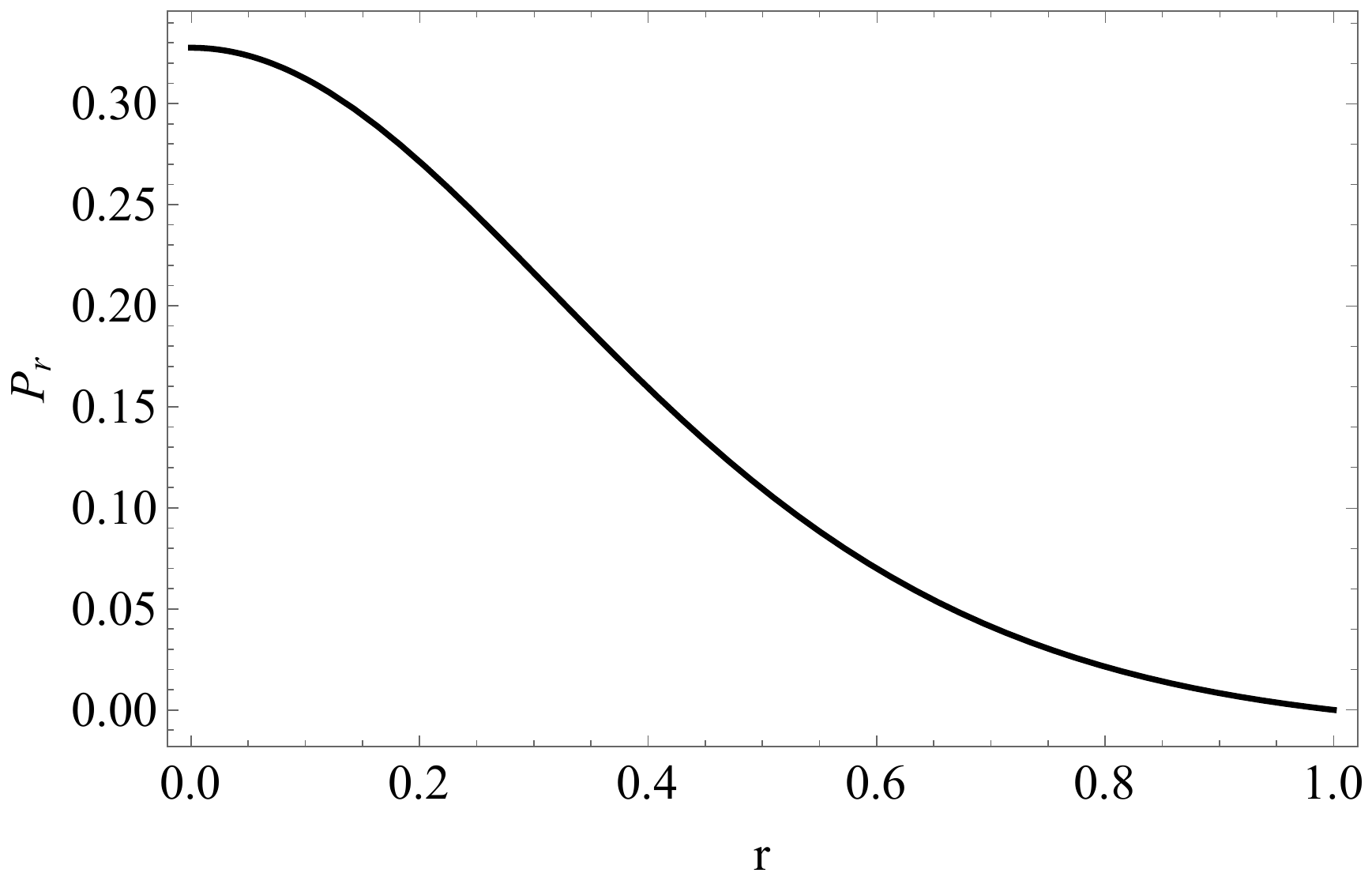}
\caption{\label{fpradial1}
$P_{r}$ as a function of the radial coordinate $r$ with $R=1$ for Durgapal IV ($q=0$) solution. 
}
\end{figure}

\begin{figure}[h!]
\centering
\includegraphics[scale=0.5]{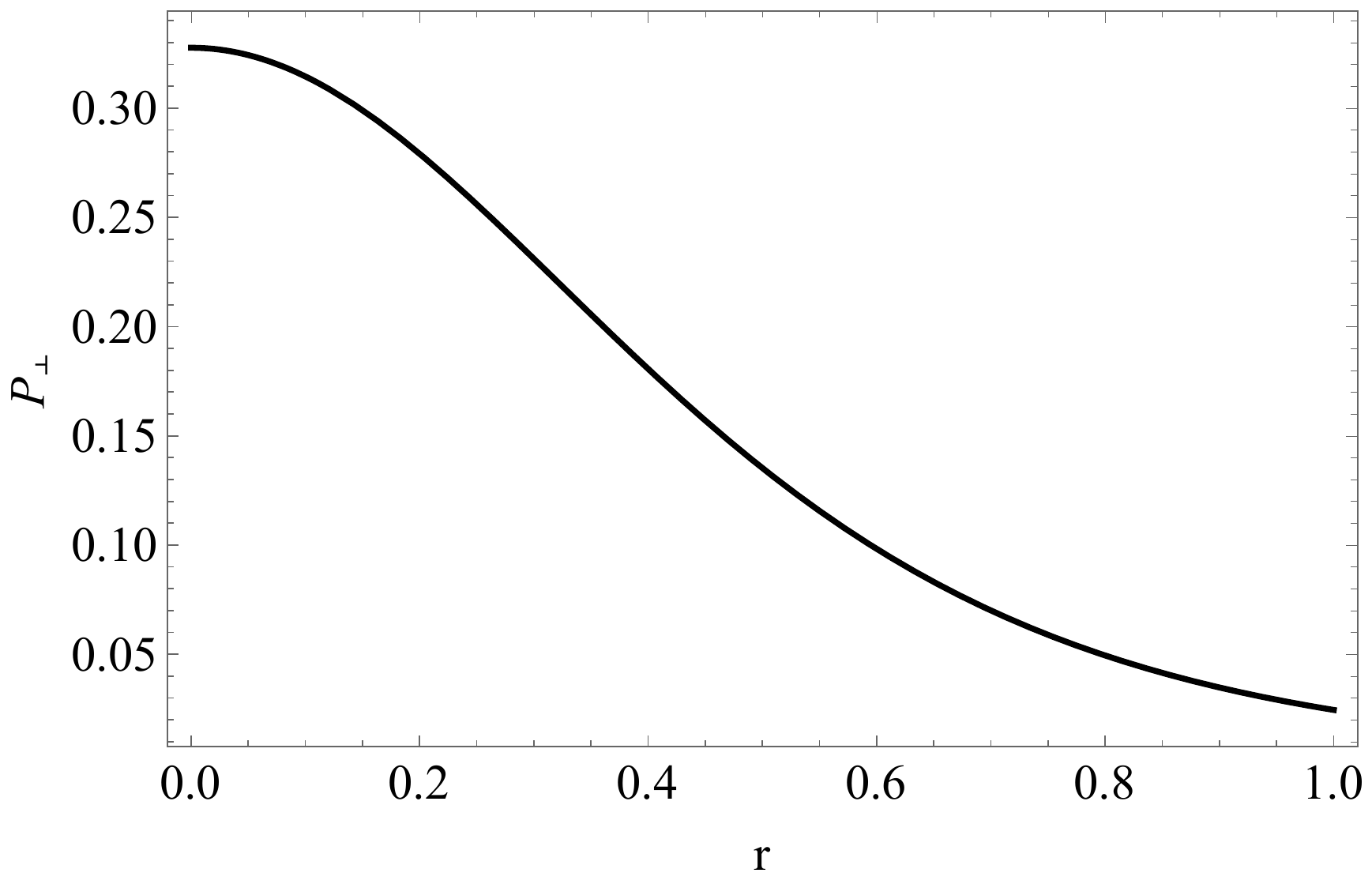}
\caption{\label{fptangencial1}
$P_{\perp}$ as a function of the radial coordinate $r$ with $R=1$ for Durgapal IV ($q=0$) solution. 
}
\end{figure}

\begin{figure}[h!]
\centering
\includegraphics[scale=0.5]{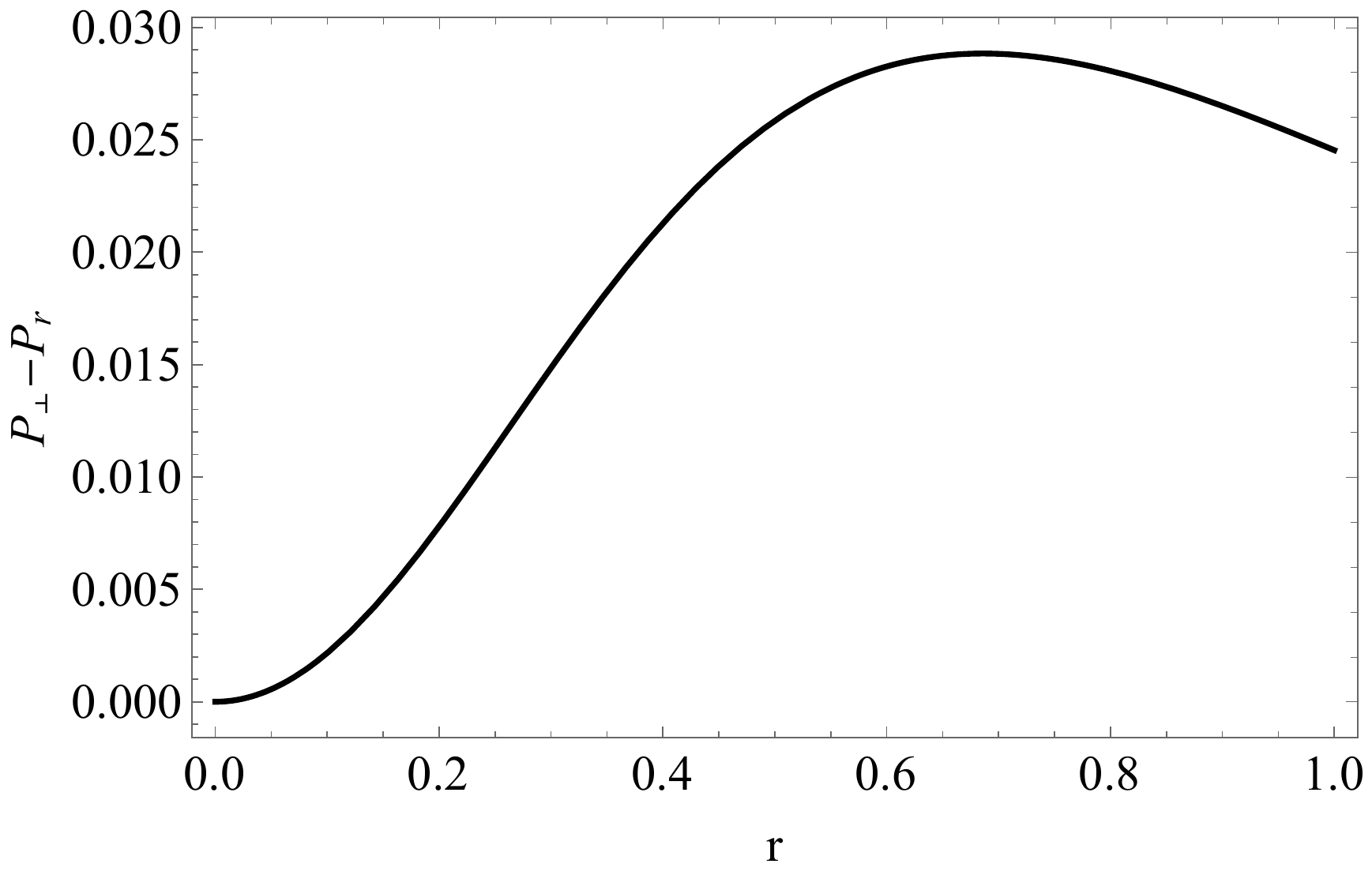}
\caption{\label{fanisotropia1}
$P_{\perp}-P_{r}$ as a function of the radial coordinate $r$ with $R=1$ for Durgapal IV ($q=0$) solution. 
}
\end{figure}

\begin{figure}[h!]
\centering
\includegraphics[scale=0.5]{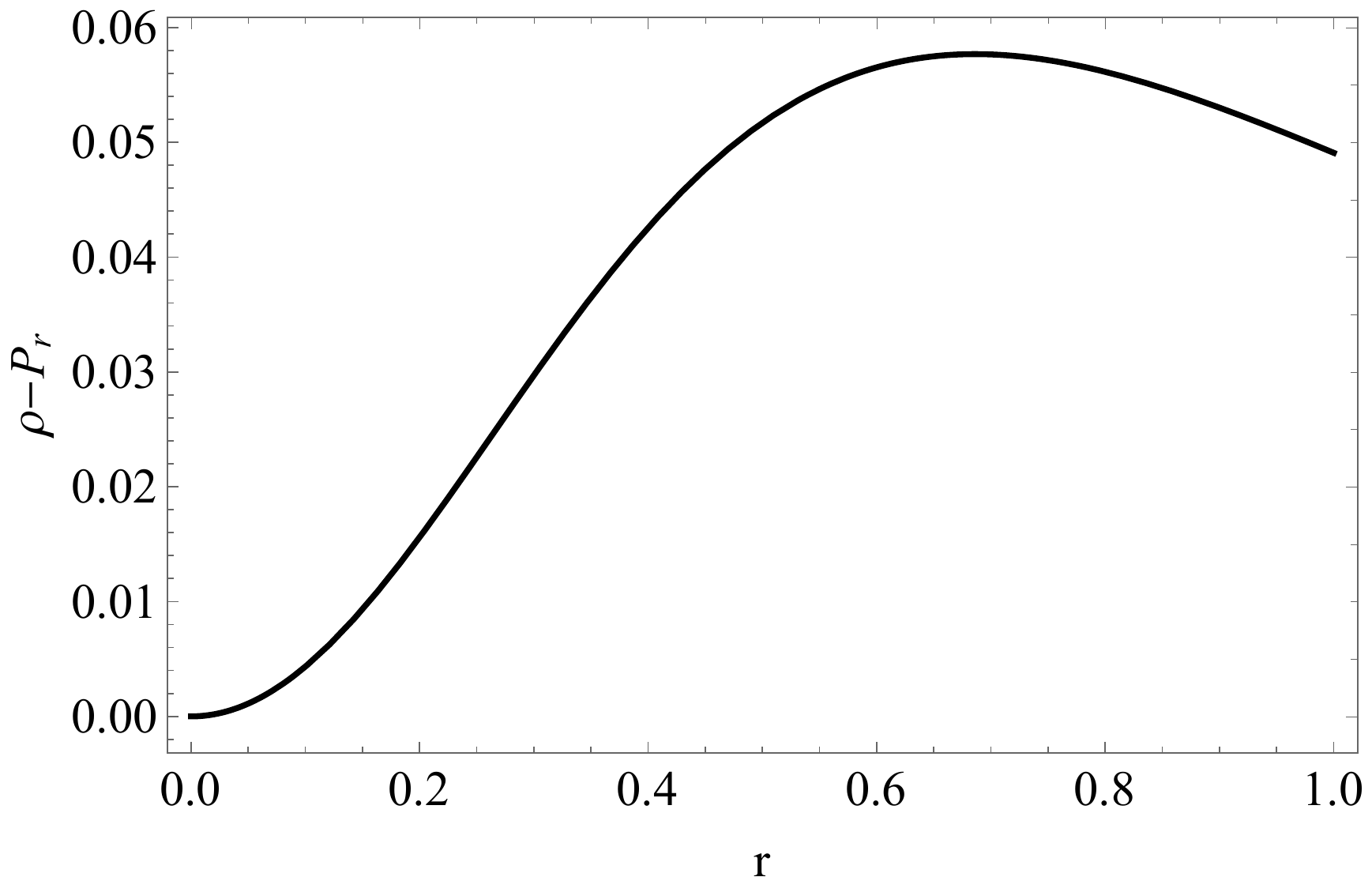}
\caption{\label{fdec1}
$\rho-P_{r}$ as a function of the radial coordinate $r$ fwith $R=1$ for Durgapal IV ($q=0$) solution. 
}
\end{figure}

\begin{figure}[h!]
\centering
\includegraphics[scale=0.5]{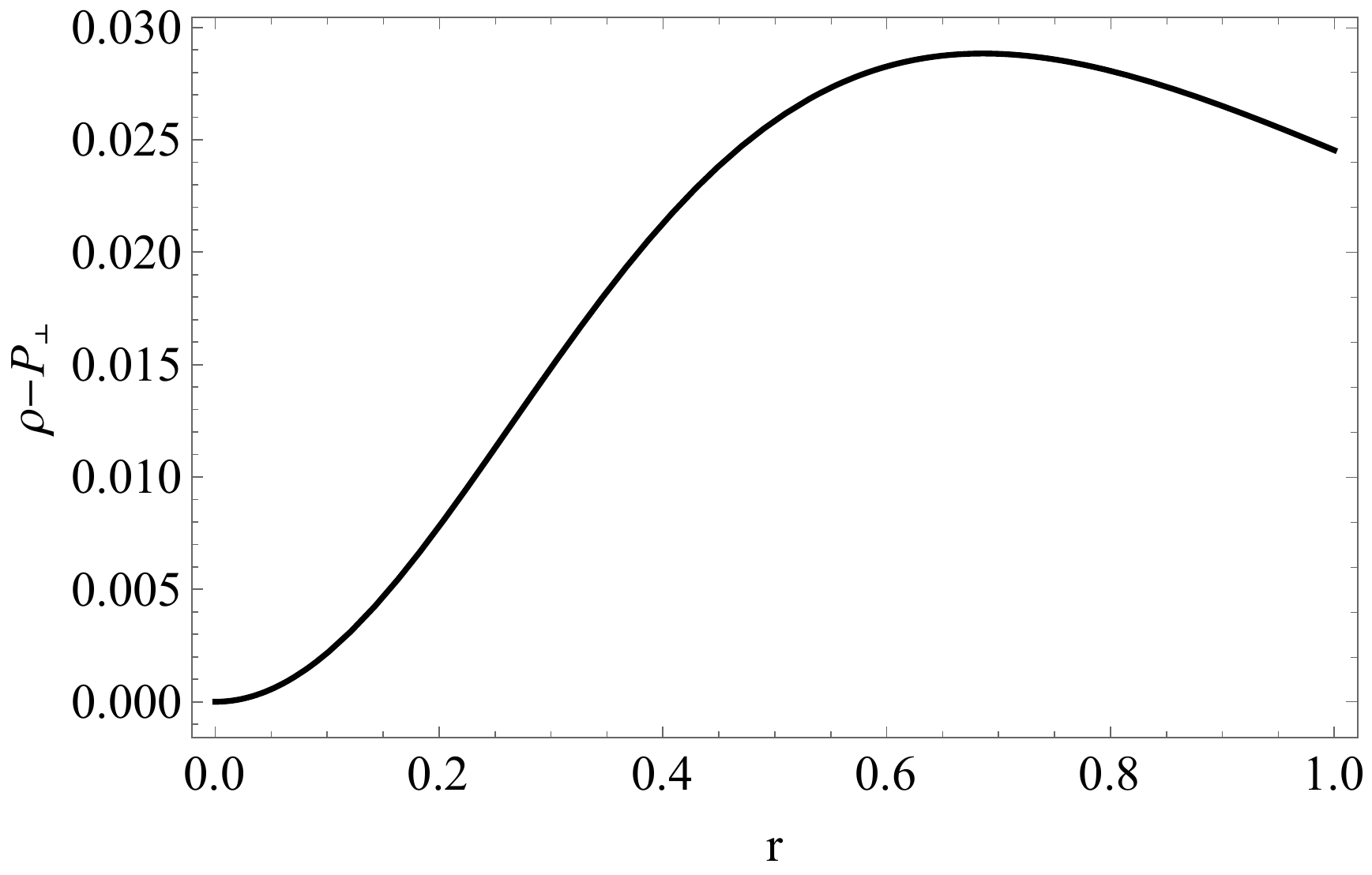}
\caption{\label{fdec21}
$\rho-P_{\perp}$ as a function of the radial coordinate $r$ with $R=1$ for Durgapal IV ($q=0$) solution. 
}
\end{figure}

Now, in figures \ref{fdensidad1}, \ref{fpradial1} and \ref{fptangencial1} we show the behavior of the matter sector (thermodynamic variables) plotted as a function of the radial coordinate $r$ using $R=1$. Note that the density is positive inside the star, reaches its maximum at the center and decreases monotonously outwards, as expected. Also, we observe that the radial pressure and tangential pressure are positive quantities inside the star, they reach their maximum at the center and then monotonously decrease outwards, representing an appropriate behavior. This, obviously translates into a suitable behavior for the anisotropy of the system as we verify by means of figure \ref{fanisotropia1}. Besides, for this model the dominant energy condition (DEC) is satisfied as it can be appreciated in both figures \ref{fdec1} and \ref{fdec21}, where the density is greater than both pressures, radial and tangential.

\section{Charged like Durgapal IV solution}

We use as a seed, in order to solve the system, the $g_{tt}$
component of the Durgapal IV solution given by,
\begin{eqnarray}\label{ch_nu_iv}
e^{\nu}=A \left(1+c r^2\right)^4,
\end{eqnarray}
but now we will consider the charged case where $q\ne 0$. As before, the vanishing complexity condition given by Eq. (\ref{YTF=0}) leads us to a relationship between the metric functions that produces
\begin{eqnarray}\label{ch_lambda_iv}
e^{\lambda}=(1+c r^{2})^{2}.
\end{eqnarray}
Replacing (\ref{ch_nu_iv}) and (\ref{ch_lambda_iv}) in (\ref{e1}), (\ref{e2}) and (\ref{e3}) we obtain
\begin{eqnarray}
\rho&=&-\frac{q^2}{8 \pi  r^4} + \frac{c \left[6 + c r^2 \left(c r^2+3\right)\right]}{8 \pi  \left(1 + c r^2\right)^3}\label{rhoq1}\\
P_{r}&=&\frac{q^2}{8 \pi  r^4} + \frac{c \left[6 - c r^2 \left(c r^2+3\right)\right]}{8 \pi  \left(1 + c r^2\right)^3}\label{prq1}\\
P_{\perp}&=&-\frac{q^2}{8 \pi  r^4}+\frac{6 c}{8 \pi  \left(1 + c r^2\right)^3}.\label{ptq1}
\end{eqnarray}

Now, let us consider the following anisotropy \cite{Dey:2020fxm,Gomez-Leyton:2020kfw}
\begin{eqnarray}\label{ch_ani_IV}
\Pi=-\alpha\frac{q^{2}}{4\pi r^{4}},
\end{eqnarray}
from where, introducing (\ref{prq1}) and (\ref{ptq1}) we arrive at
\begin{eqnarray}\label{carga1}
q=\frac{c r^3 \sqrt{3 + c r^2}}{\sqrt{2(1 + \alpha)} \left(1 + c r^2\right)^{3/2}}.
\end{eqnarray}
The matching conditions with the outside metric leads to
\begin{eqnarray}
A \left(c R^2+1\right)^4&=&-\frac{2 M}{R}+\frac{Q^2}{R^2}+1\\
\frac{1}{\left(c R^2+1\right)^2}&=&-\frac{2 M}{R}+\frac{Q^2}{R^2}+1\\
\frac{Q^2}{R^4}&=&\frac{c \left(c R^2 \left(c R^2+3\right)-6\right)}{\left(c R^2+1\right)^3}\\
Q^2&=&\frac{c^2 R^6 \left(c R^2+3\right)}{2 (\alpha +1) \left(c R^2+1\right)^3},
\end{eqnarray}
from where we get the expressions
\begin{eqnarray}
\frac{Q}{\xi_{1}}&=&\bigg(3 \sqrt{3}R^2 (2 \alpha +1) \sqrt{(2 \alpha +1) (22 \alpha +19) }\nonumber\\
&&-9 (4 \alpha  (37 \alpha +67)+121) R^2\bigg)^{1/2}\\
c&=&\frac{\sqrt{3} \sqrt{(2 \alpha +1) (22 \alpha +19) }}
{(2 \alpha +1)2 R^2}-\frac{3}{2 R^2} ,\\  
\frac{A}{\xi_{2}}&=& 1-\frac{2 M}{R}+\frac{9 (4 \alpha  (37 \alpha +67)+121) }{8 (8 \alpha +7)^3 }\nonumber\\
&&-
\frac{3 \sqrt{3} (2 \alpha +1) \sqrt{(2 \alpha +1) (22 \alpha +19) }}{8 (8 \alpha +7)^3 } , \\
\frac{M}{\xi_{3}}&=&\frac{3 (10 \alpha +9) R^2-\xi_{4}}{ (34 \alpha +29) R^2-\xi_{4}},
\end{eqnarray}
whith
\begin{eqnarray}
\xi_{1}&=&\frac{1}{2 \sqrt{2} \sqrt{-(8 \alpha +7)^3}}\\
\xi_{2}&=&16\bigg(\frac{\sqrt{3} (22 \alpha +19)}{\sqrt{(2 \alpha +1) (22 \alpha +19) }}-1\bigg)^{-4}\\
\xi_{3}&=&\frac{(4 \alpha +5) R}{(8 \alpha +7)} \\
\xi_{4}&=&\sqrt{3}R^{2} \sqrt{(2 \alpha +1) (22 \alpha +19)}.
\end{eqnarray}

Due to the choice of the specific function for the local anisotropy, which is proportional to the electric charge, increasing the charge (or field) value is equivalent to increasing the anisotropy of the system (note that it can also be modulated with the $\alpha$ parameter) as seen from equation (\ref{ch_ani_IV}). In addition, we observe from equations (\ref{rhoq1}), (\ref{prq1}) and (\ref{ptq1}) that, in effect, one could choose effective variables for the fluid in the sense that they incorporate the term associated with the electric field, which will be translated into the effective anisotropy of the system.

\begin{figure}[h!]
\centering
\includegraphics[scale=0.5]{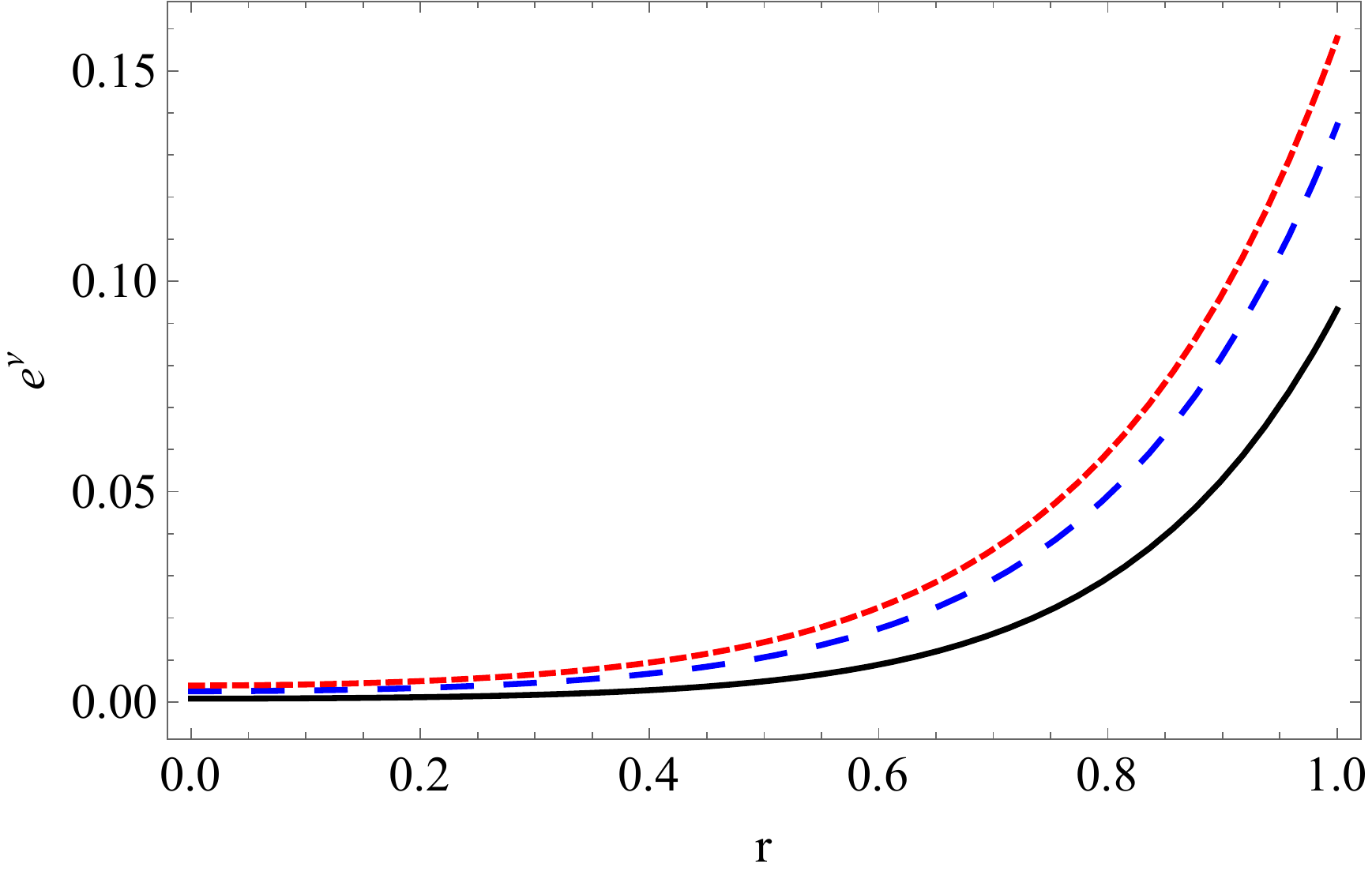}
\caption{\label{ch_enu_iv}
Metric $e^{\nu}$ as a function of the radial coordinate $r$ for the Durgapal IV charged ($q\ne 0$) model. $R=1$ and $\alpha=0$ (black line), 
$\alpha=1$ (blue dashed line) and $\alpha=3$ (red dotted line). 
}
\end{figure}

\begin{figure}[h!]
\centering
\includegraphics[scale=0.5]{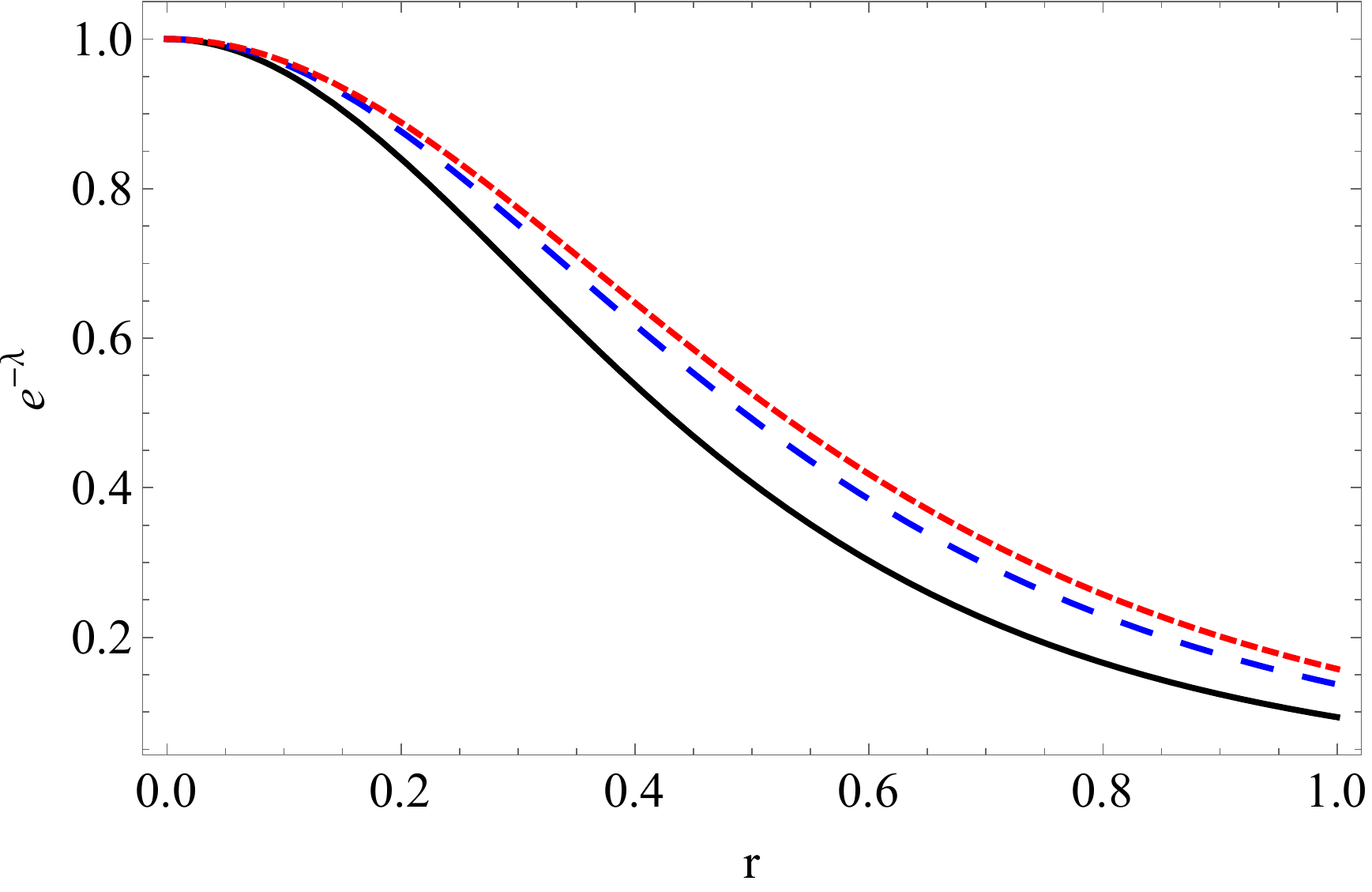}
\caption{\label{ch_elambda_iv}
Metric $e^{-\lambda}$ as a function of the radial coordinate $r$ for the Durgapal IV charged ($q\ne 0$) model. $R=1$ and $\alpha=0$ (black line), 
$\alpha=1$ (blue dashed line) and $\alpha=3$ (red dotted line). 
}
\end{figure}

In figures \ref{ch_enu_iv} and \ref{ch_elambda_iv} we show the behavior of the metric functions, with respect to the radial coordinate $r$ for the Durgapal IV charged ($q\ne 0$) model. We observe that these functions are positive, finite and free of singularities, as they should be for a physical accepted solution. Furthermore, evaluated at zero, both reach the expected values, $e^{-\lambda(0)}=1$ and $e^{\nu(0)}=const$.

\begin{figure}[h!]
\centering
\includegraphics[scale=0.5]{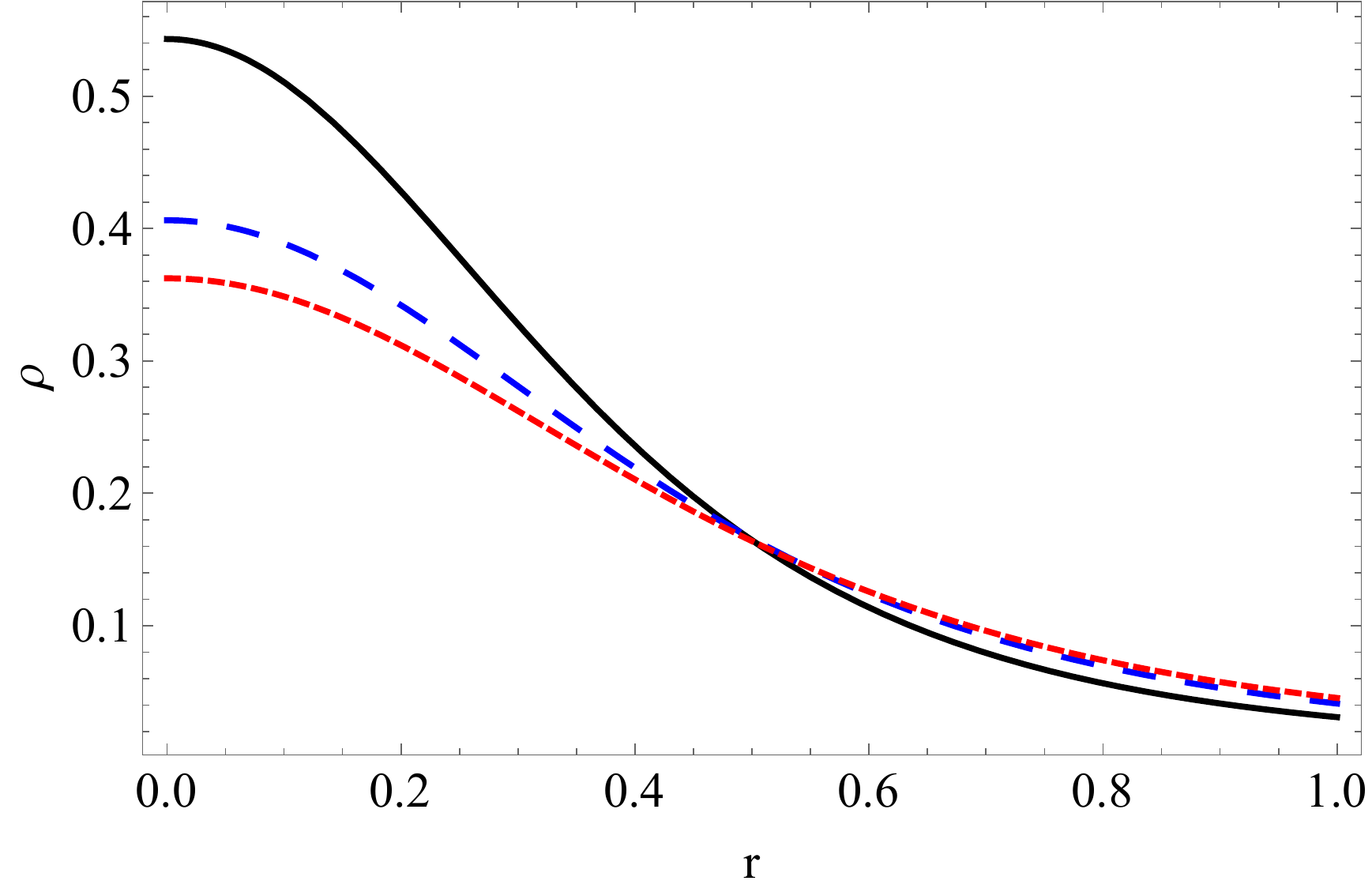}
\caption{\label{ch_densidad_iv}
$\rho$ as a function of the radial coordinate $r$ for the Durgapal IV charged ($q\ne 0$) model. $R=1$ and $\alpha=0$ (black line), 
$\alpha=1$ (blue dashed line) and $\alpha=3$ (red dotted line). 
}
\end{figure}

\begin{figure}[h!]
\centering
\includegraphics[scale=0.5]{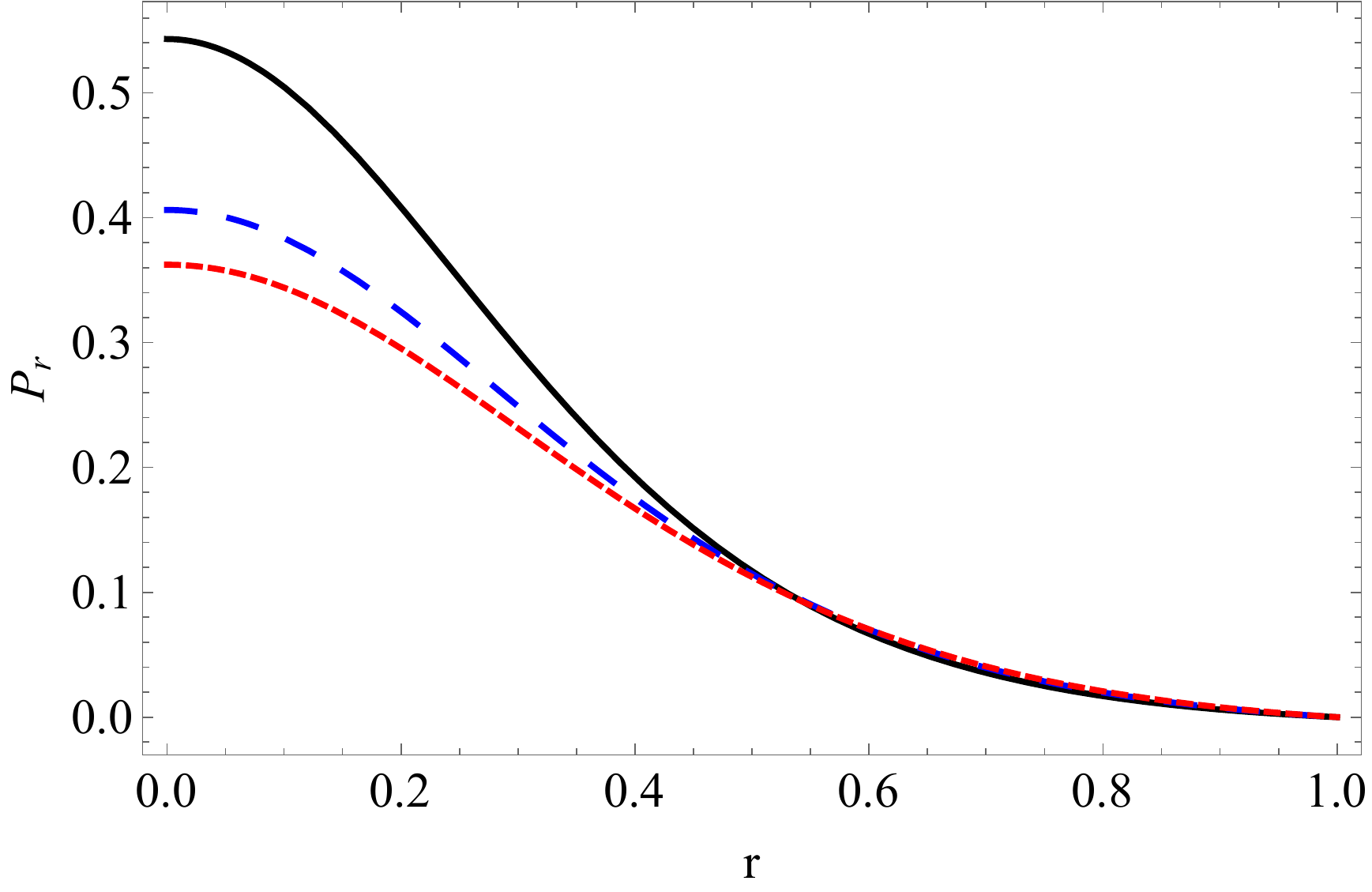}
\caption{\label{ch_pradial_iv}
$P_{r}$ as a function of the radial coordinate $r$ for the Durgapal IV charged ($q\ne 0$) model. $R=1$ and $\alpha=0$ (black line), 
$\alpha=1$ (blue dashed line) and $\alpha=3$ (red dotted line). 
}
\end{figure}

\begin{figure}[h!]
\centering
\includegraphics[scale=0.5]{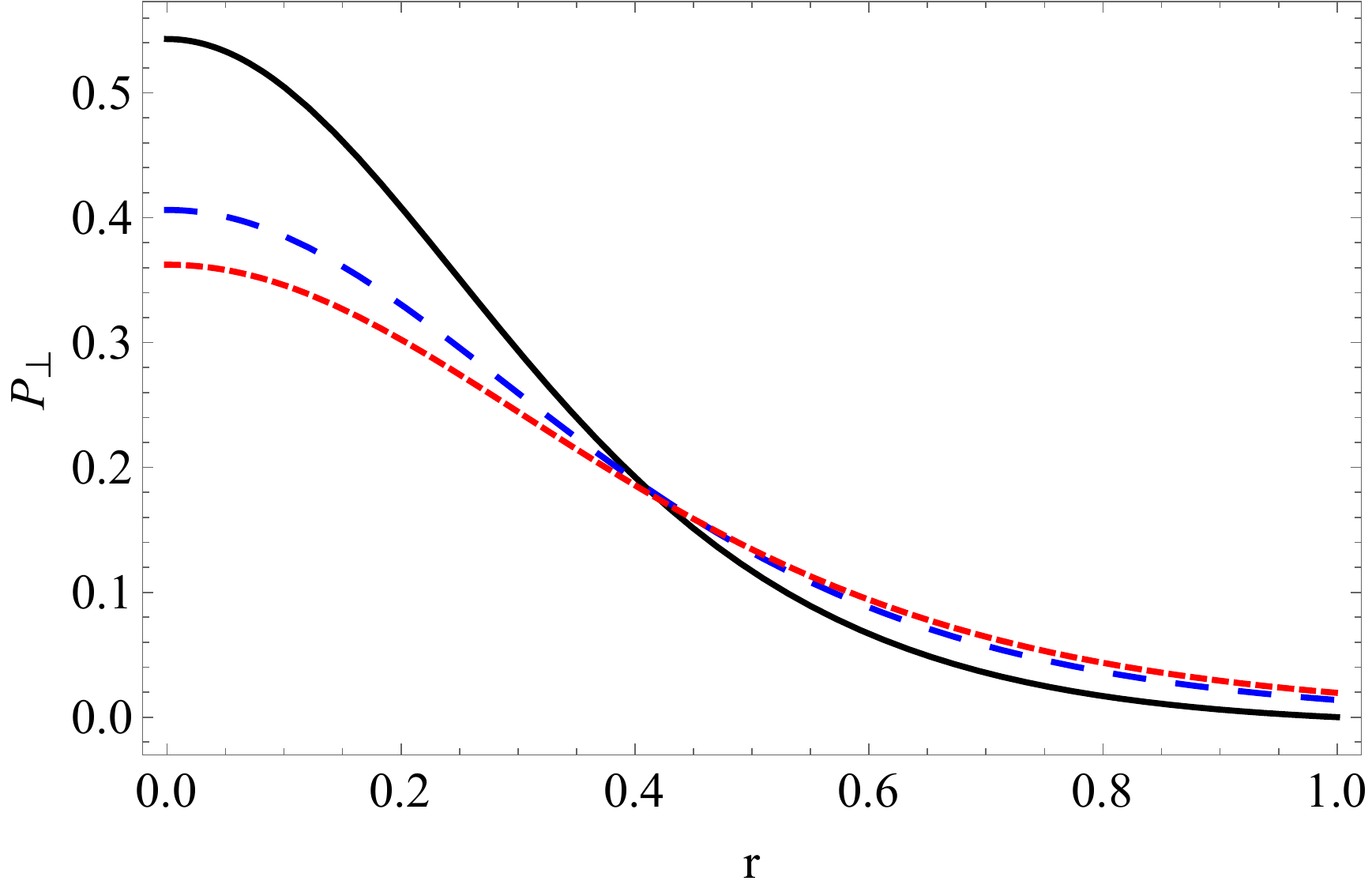}
\caption{\label{ch_ptangencial_iv}
$P_{\perp}$ as a function of the radial coordinate $r$ for the Durgapal IV charged ($q\ne 0$) model. $R=1$ and $\alpha=0$ (black line), 
$\alpha=1$ (blue dashed line) and $\alpha=3$ (red dotted line). 
}
\end{figure}

\begin{figure}[h!]
\centering
\includegraphics[scale=0.5]{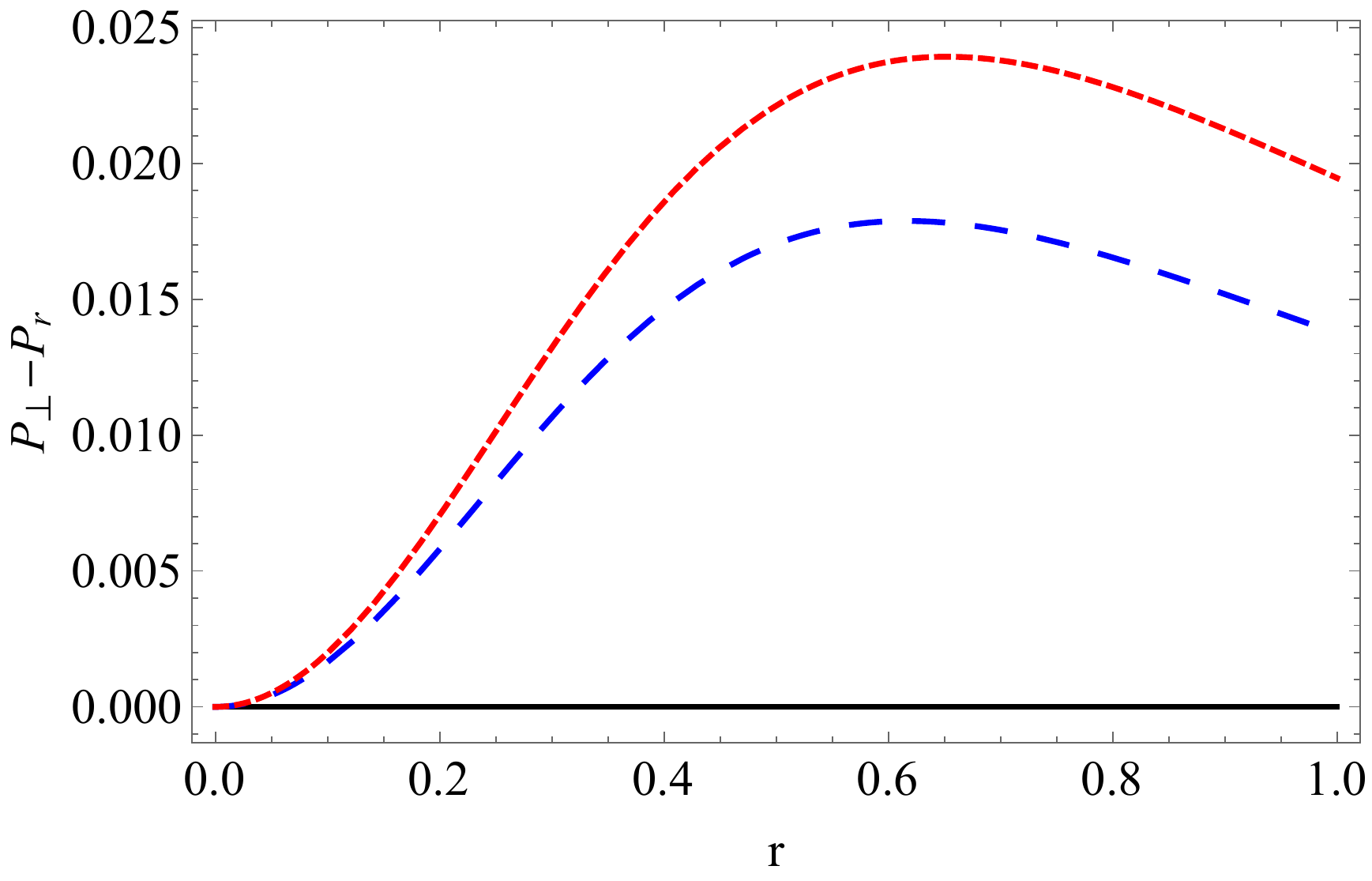}
\caption{\label{ch_anisotropia_iv}
$P_{\perp}-P_{r}$ as a function of the radial coordinate $r$ for the Durgapal IV charged ($q\ne 0$) model. $R=1$ and $\alpha=0$ (black line), 
$\alpha=1$ (blue dashed line) and $\alpha=3$ (red dotted line). 
}
\end{figure}

\begin{figure}[h!]
\centering
\includegraphics[scale=0.5]{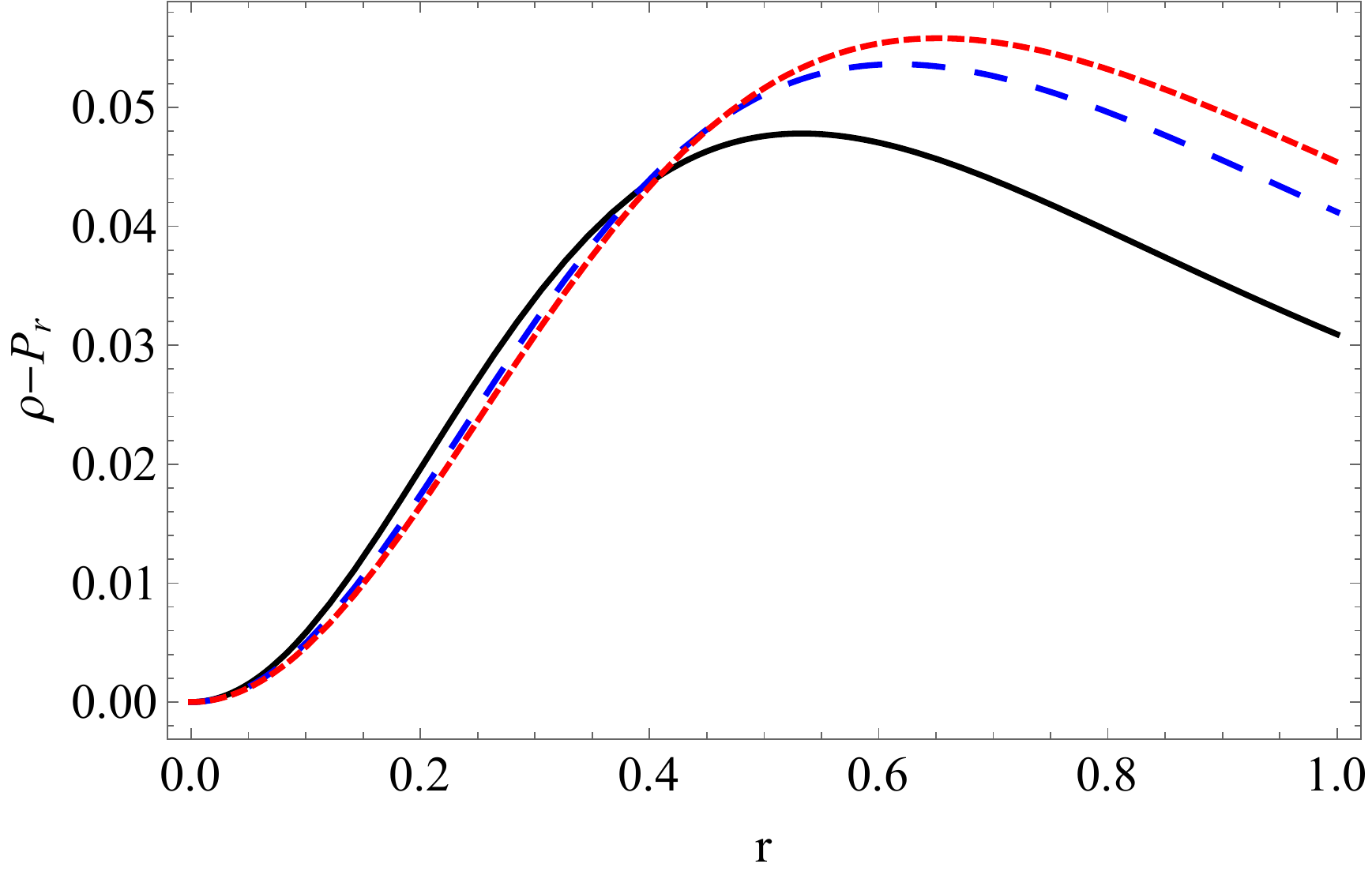}
\caption{\label{ch_dec1_iv}
$\rho-P_{r}$ as a function of the radial coordinate $r$ for the Durgapal IV charged ($q\ne 0$) model. $R=1$ and $\alpha=0$ (black line), 
$\alpha=1$ (blue dashed line) and $\alpha=3$ (red dotted line). 
}
\end{figure}

\begin{figure}[h!]
\centering
\includegraphics[scale=0.5]{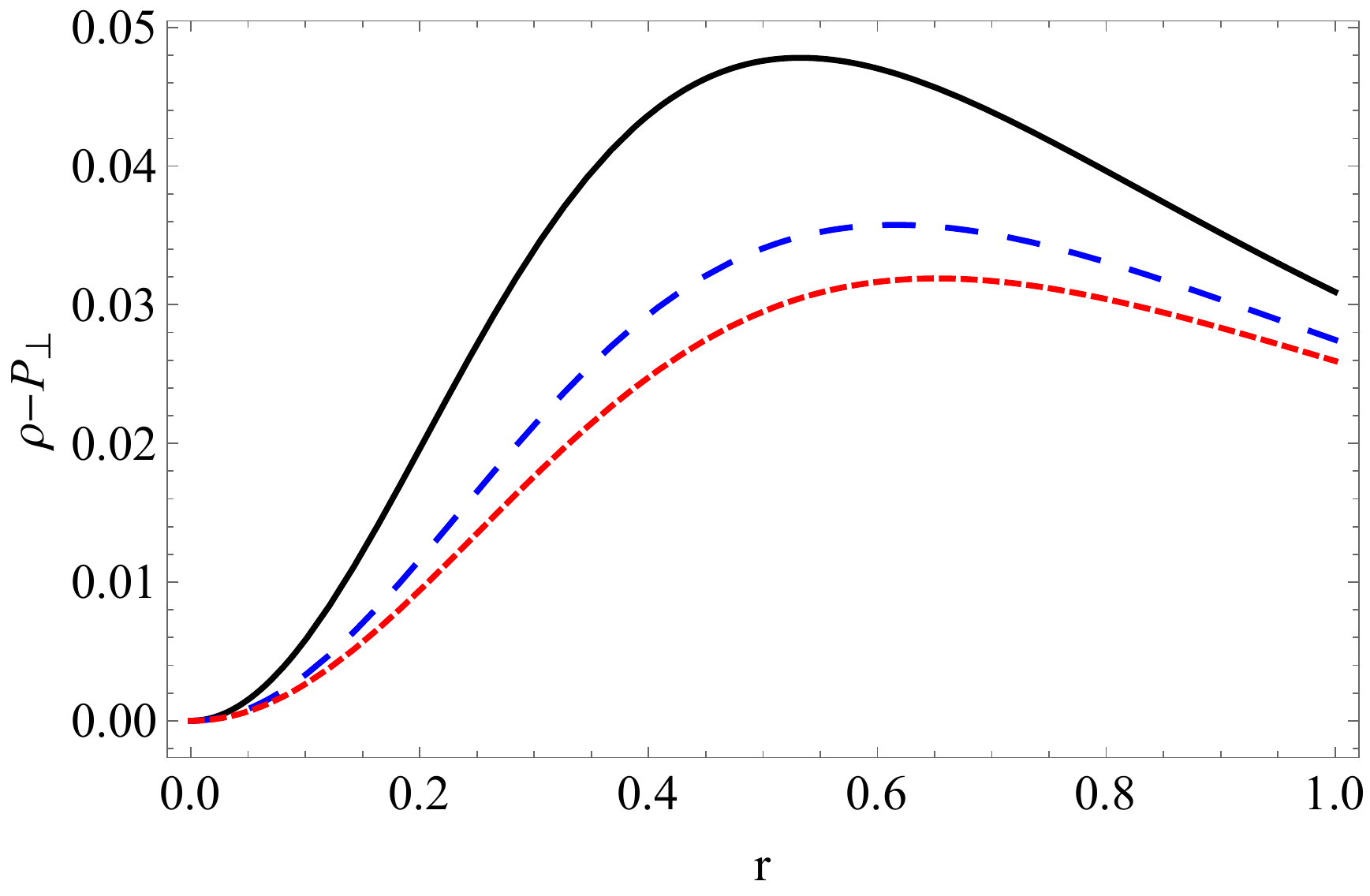}
\caption{\label{ch_dec2_iv}
$\rho-P_{\perp}$ as a function of the radial coordinate $r$ for the Durgapal IV charged ($q\ne 0$) model. $R=1$ and $\alpha=0$ (black line), 
$\alpha=1$ (blue dashed line) and $\alpha=3$ (red dotted line). 
}
\end{figure}

In figures \ref{ch_densidad_iv}, \ref{ch_pradial_iv} and \ref{ch_ptangencial_iv} the matter sector and the fluid tensions are shown. Note that the density and pressure (radial and tangential) behave as expected: positive quantities inside the star, their maximum is at the centre and then decrease monotonously outwards. As a consequence of this, the anisotropy shown in the figure \ref{ch_anisotropia_iv} will have the appropriate behavior. For this model we note that the $\alpha$ parameter controls the anisotropy present in the compact object, so that $\alpha =0$ represents the isotropic charged case. Also, setting $q=0$ automatically returns us to the previous anisotropic un-charged Durgapal IV model. In the internal regions (up to values for the radial coordinate between $0.4-0.6$) of the stellar object, all these variables decrease with the increasing anisotropy, as shown in the figure \ref{ch_anisotropia_iv}. For the same interval of values of $r$ the curve of the anisotropy function is smoothed (reaches a maximum approximately for $r=0.6$) so from there on the behavior of the matter sector is the opposite. Also, for this charged Durgapal IV anisotropic model, the energy condition (DEC) is also satisfied since $\rho$ $\geq$ $P_{r}$, and $\rho$ $\geq$ $P_{\perp}$, as can be seen in the figures \ref{ch_dec1_iv} and \ref{ch_dec2_iv}. 

\begin{figure}[h!]
\centering
\includegraphics[scale=0.5]{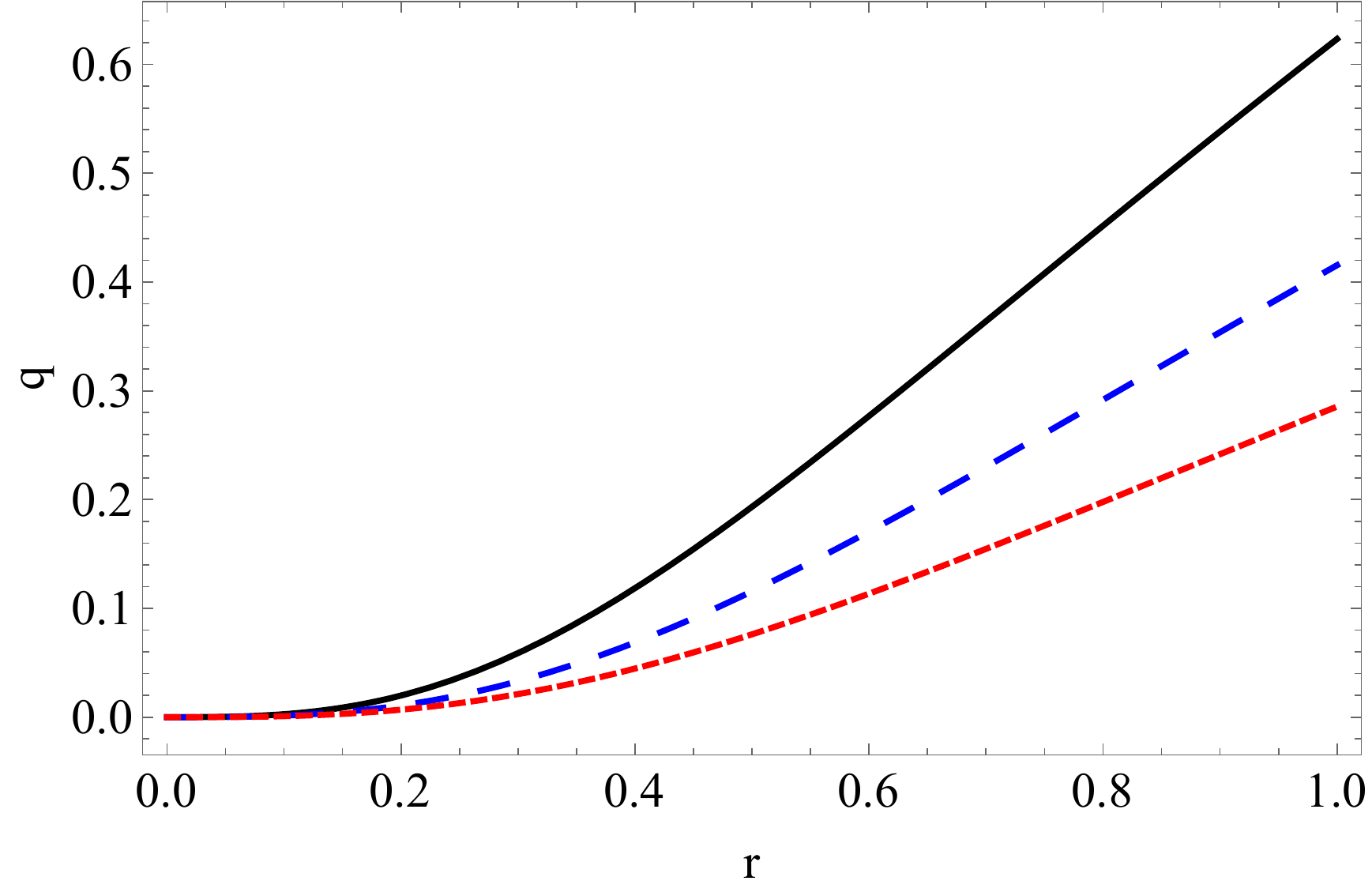}
\caption{\label{ch_carga_iv}
$q$ as a function of the radial coordinate $r$ for the Durgapal IV charged ($q\ne 0$) model. $R=1$ and $\alpha=0$ (black line), 
$\alpha=1$ (blue dashed line) and $\alpha=3$ (red dotted line). 
}
\end{figure}

Finally we show in figure \ref{ch_carga_iv} the behavior of the interior electric charge (the electric field will be proportional to the charge) for this model. Initially, in the center Gauss's law tells us that it must be zero and then grows monotonously to its full value at the surface $q=q_{\Sigma}=Q$, as it should be. Although the behavior of the interior charge is linked to the anisotropy of the system, we can observe this fact from the expression (\ref{ch_ani_IV}), the existence of charge inside the stellar object is independent of whether or not there is anisotropy present, as we can see for the black line ($\alpha =0$) of the same figure. The internal charge decreases, throughout the range of values of $r$ by increasing $\alpha$ and therefore the local anisotropy. This could have been read directly from the equation (\ref{carga1}).

\section{Durgapal V solution using vanishing complexity}

We now proceed completely analogously with the Durgapal V model, both in the uncharged ($q=0$) and charged ($q\ne 0$) cases. The objective of carrying out this other case is to be able to have a comparative framework with the previous results and verify their consistency. In addition, for completeness, we present a self-contained work. First we consider the uncharged Durgapal V solution. Setting $q=0$ in Eqs. \eqref{e1}, \eqref{e2} and \eqref{e3} we use the $g_{tt}$ metric coefficient as a seed (a well-behaved known solution) that will allow us to obtain the complete solution of the problem considered. So, we start with the metric temporal function,
\begin{eqnarray}\label{nu}
e^{\nu}=A \left(1+c r^2\right)^5\,.
\end{eqnarray}
The vanishing complexity condition \eqref{YTF=0}, as before, translates into a relationship between the metric variables and enables us to find the radial metric coefficient, given by
\begin{eqnarray}\label{lambda}
e^{\lambda}=(1+c r^{2})^{3}\,.
\end{eqnarray}
Replacing both metric coefficients in Eqs. \eqref{e1}, \eqref{e2} and \eqref{e3} (with $q=0$) we find
\begin{eqnarray}
\rho &=& \frac{c \left(9 + 6cr^2 + 4c^2r^4 + c^3r^6\right)}{8 \pi  \left(1 + c r^2\right)^4}\;,\\
P_{r} &=& \frac{c \left( 7 - 6cr^2 - 4c^2r^4 - c^3r^6 \right)}{8 \pi  \left(1 + c r^2\right)^4},\\
P_{\perp} &=& \frac{7 c}{8 \pi  \left(1 + c r^2\right)^4}\,.
\end{eqnarray}
We note that in the previous equations we only have one free parameter, the constant $c$. Also, using the previous equations (for $P_r$ and $P_{\perp}$) we arrive to the local anisotropy function of the solution,
\begin{equation}
    \Delta = \frac{c^2 r^2(6 + 4cr^2 + c^2r^4)}{8\pi (1+cr^2)^4}\,.
\end{equation}
Now, the matching conditions \eqref{MC01}, \eqref{MC02} and \eqref{MC03}, with $Q=0$ (which corresponds to the Schwarzschild vacuum solution), allow us to obtain the following expressions
\begin{eqnarray}
A &=& \frac{1}{(1 - cR^2)^8}\;,\\
M &=& \frac{R}{2} \left[ 1 - \frac{1}{(1 + cR^2)^3} \right] \;,\\
cR^2 &=& \frac{1}{3} \Bigg\{
\left[ \frac{1}{2}\bigg(277 + 3\sqrt{8529}\bigg)\right]^{1/3} \nonumber\\
& & - 2 \left( \frac{2}{277 + 3\sqrt{8529}} \right)^{1/3} - 4
\Bigg\}  \;.
\end{eqnarray}
Note that only the  parameter $R$ remains free, so as before (un--charged Durgapal IV) this model presents the peculiarity of not having free parameters that allows controlling the anisotropy of the system. Also, note that for this model the compactness parameter ($M/R$) is 0.4047 that is close to the compactness obtained in the previous un-charged Durgapal IV model.

\begin{figure}[h!]
\centering
\includegraphics[scale=0.38]{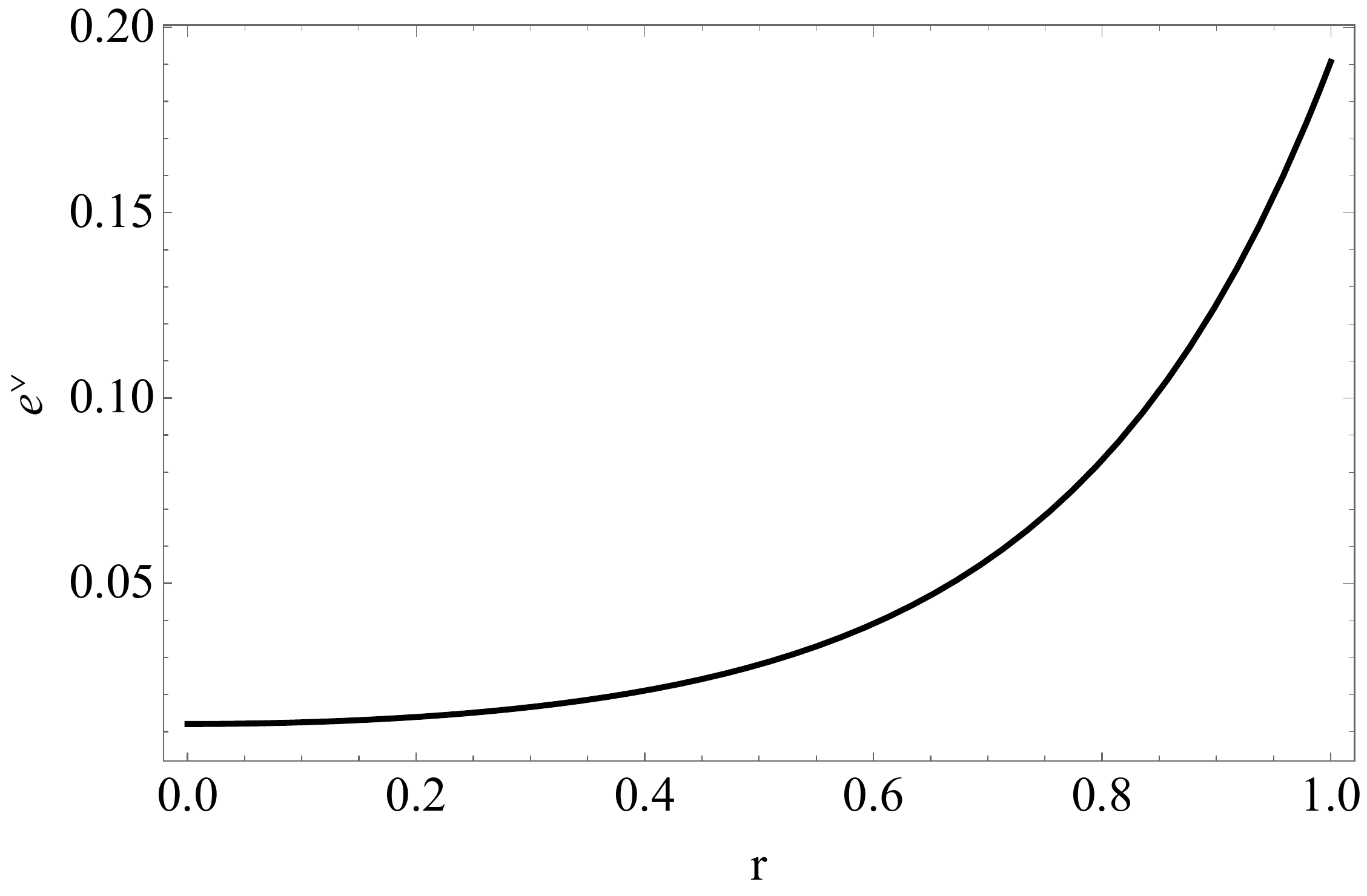}
\caption{\label{enu2}
Metric temporal component $e^{\nu}$ as a function of the radial coordinate $r$ for the un--charged Durgapal V (vanishing complexity) solution with $R=1$.
}
\end{figure}

\begin{figure}[h!]
\centering
\includegraphics[scale=0.38]{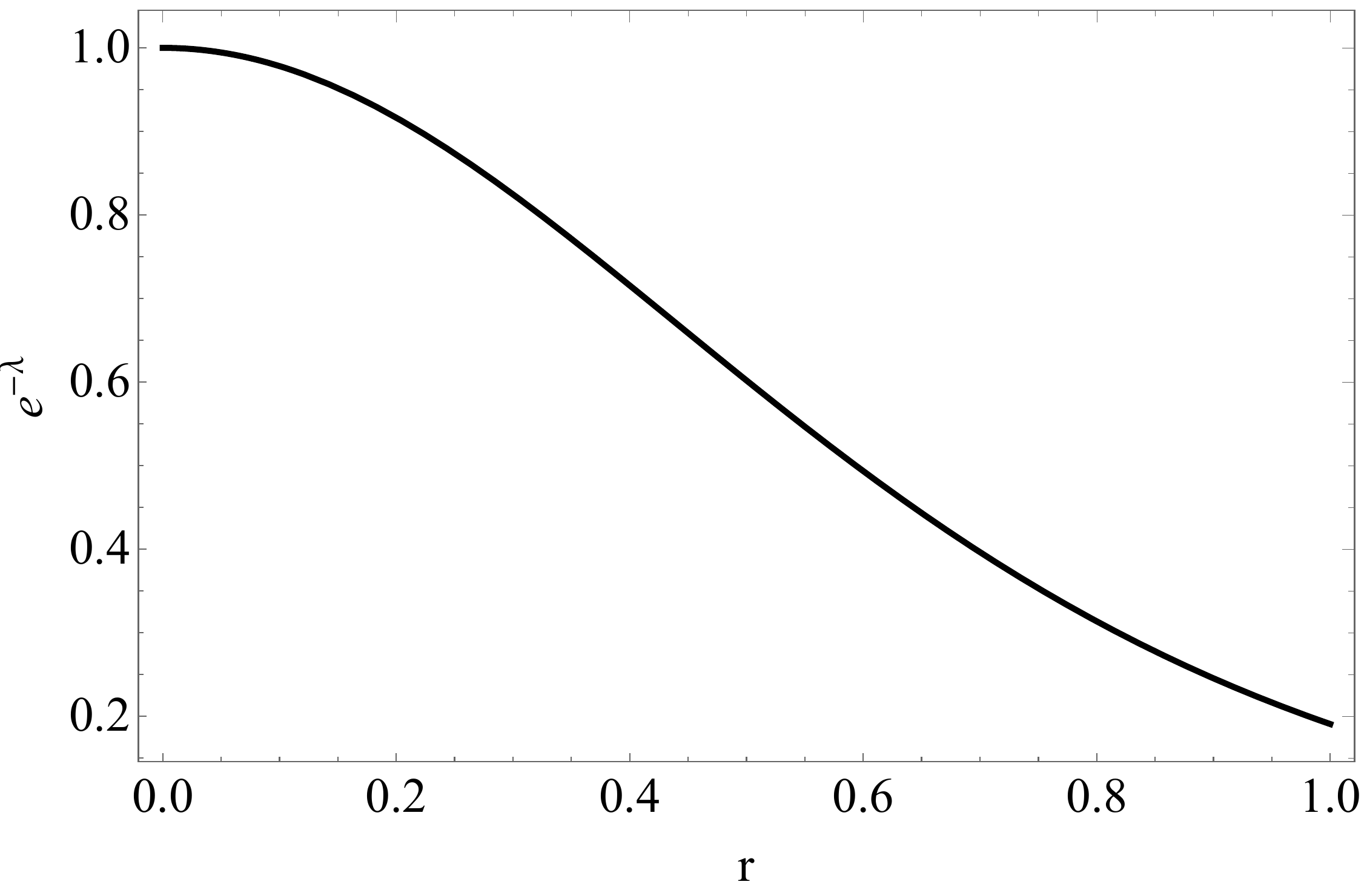}
\caption{\label{elambda2}
Metric radial component $e^{-\lambda}$ as a function of the radial coordinate $r$ for the un--charged Durgapal V (vanishing complexity) solution with $R=1$. 
}
\end{figure}

In figures \ref{enu2} and \ref{elambda2} we show the graphic behavior of the temporal and radial metric functions respectively, plotted as a function of $r$ for the un--charged Durgapal V (vanishing complexity) solution with $R=1$. Note that both metric functions present the appropriate behavior as it occurred in the case of the Durgapal IV. The temporal metric function decreases while the radial increases as long as the radius increases up to its value on the surface, as expected.

\begin{figure}[h!]
\centering
\includegraphics[scale=0.38]{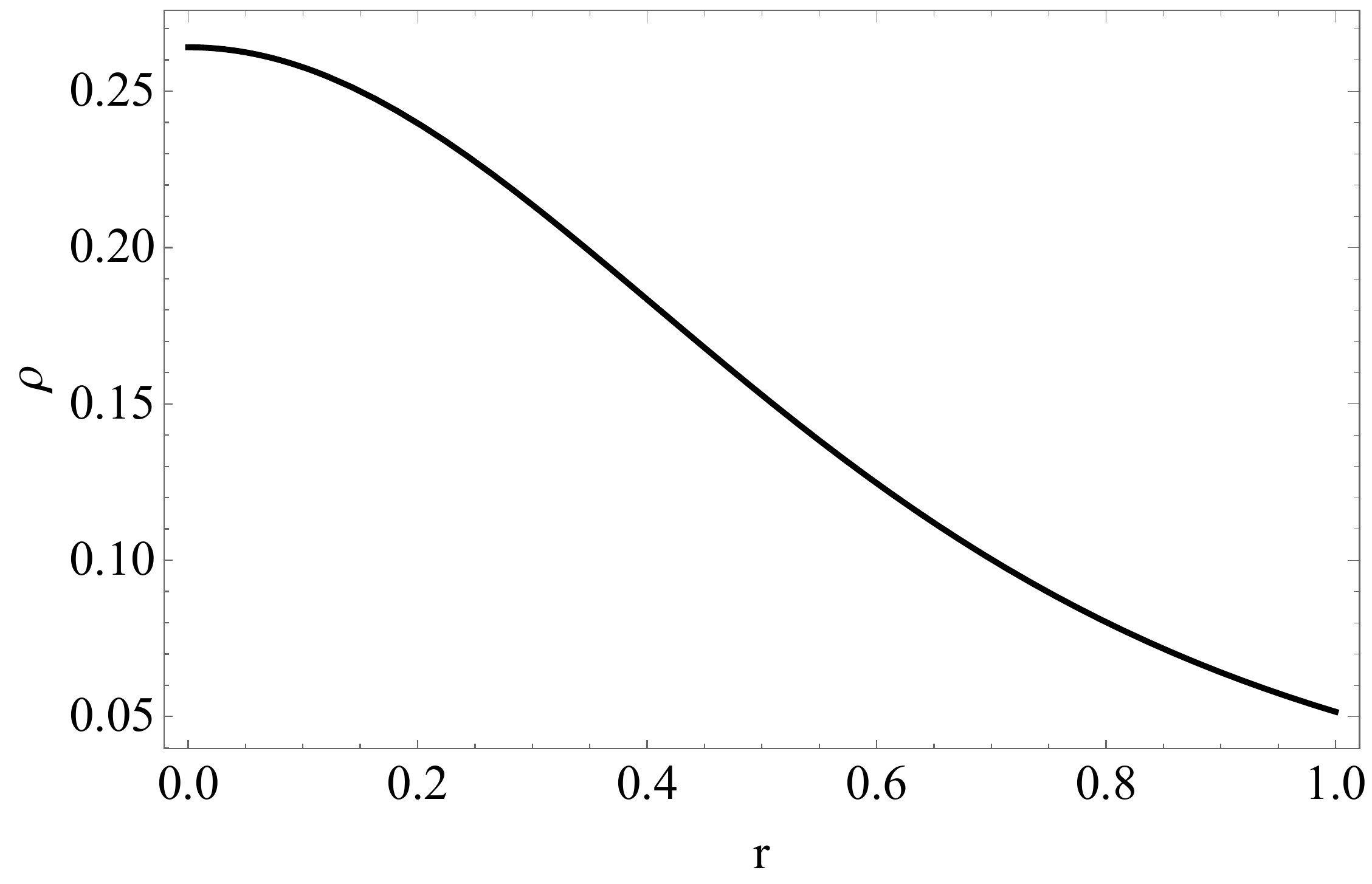}
\caption{\label{densidad2}
$\rho$ as a function of the radial coordinate $r$ for the un--charged Durgapal V (vanishing complexity) solution with $R=1$.
}
\end{figure}

\begin{figure}[h!]
\centering
\includegraphics[scale=0.38]{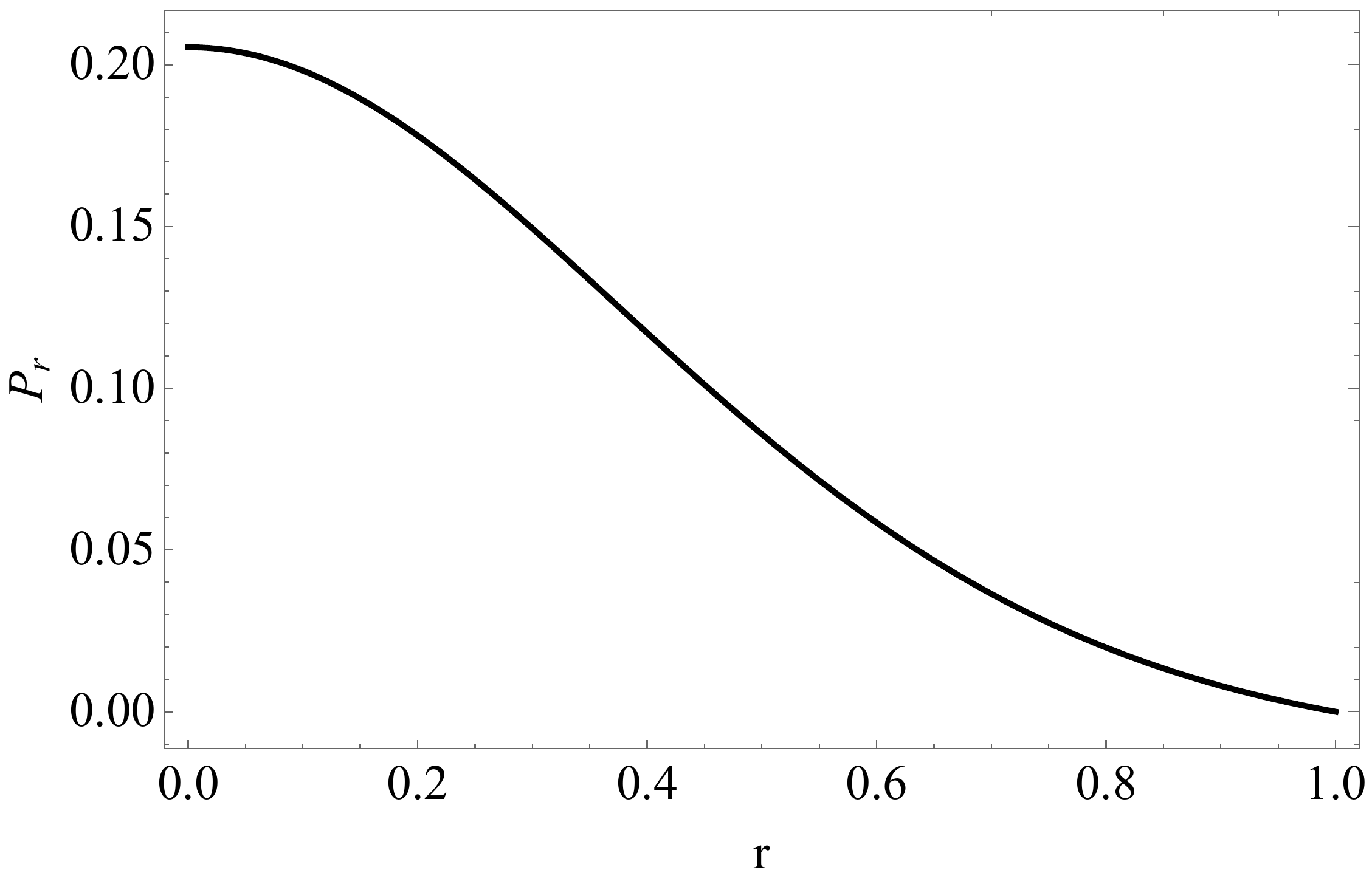}
\caption{\label{pradial2}
$P_{r}$ as a function of the radial coordinate $r$ for the un--charged Durgapal V (vanishing complexity) solution with $R=1$. 
}
\end{figure}

\begin{figure}[h!]
\centering
\includegraphics[scale=0.38]{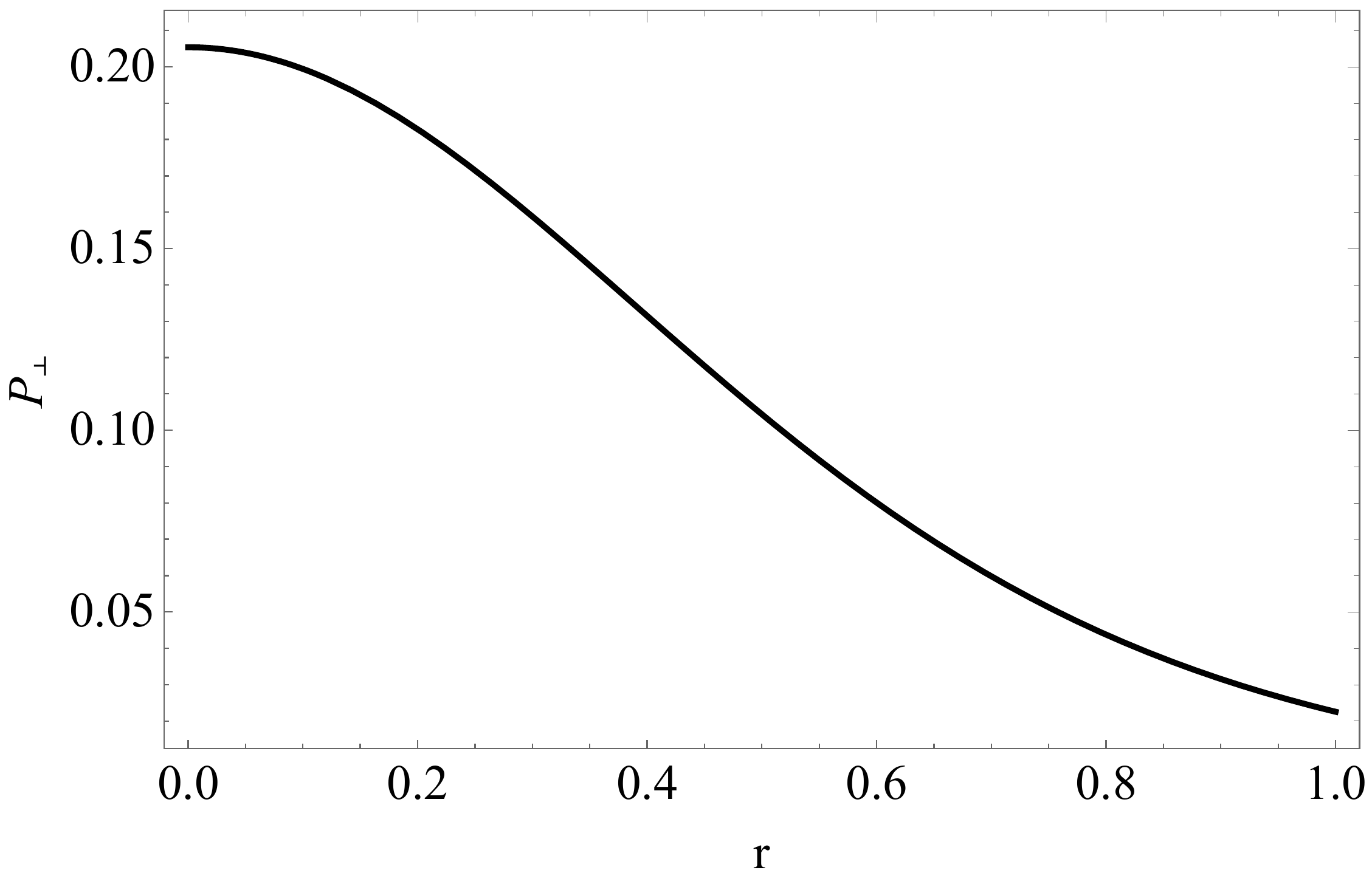}
\caption{\label{ptangencial2}
$P_{\perp}$ as a function of the radial coordinate $r$ for the un--charged Durgapal V (vanishing complexity) solution with $R=1$. 
}
\end{figure}

\begin{figure}[h!]
\centering
\includegraphics[scale=0.38]{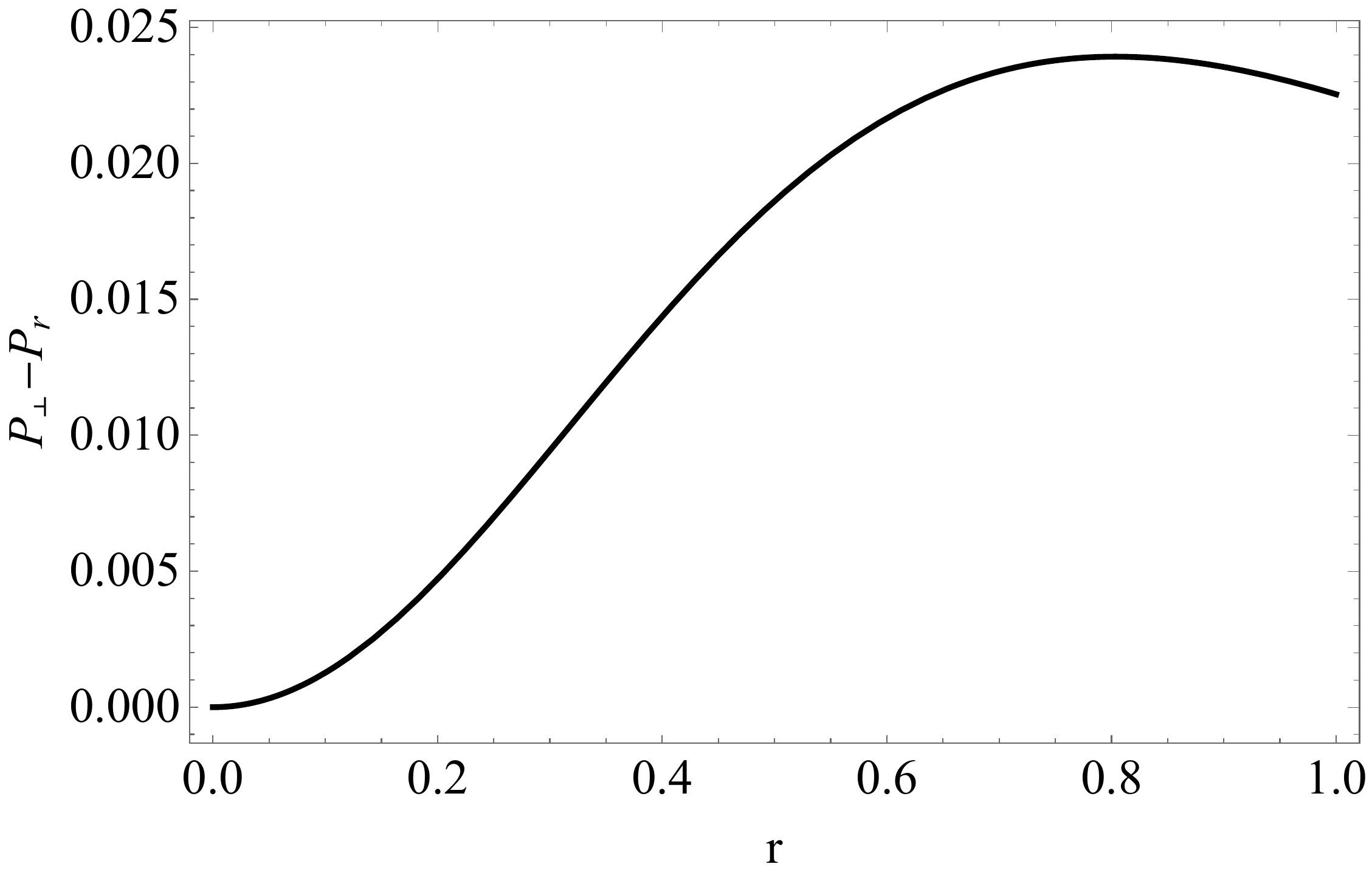}
\caption{\label{anisotropia2}
$\Delta = P_{\perp}-P_{r}$ as a function of the radial coordinate $r$ for the un--charged Durgapal V (vanishing complexity) solution with $R=1$. 
}
\end{figure}

\begin{figure}[h!]
\centering
\includegraphics[scale=0.38]{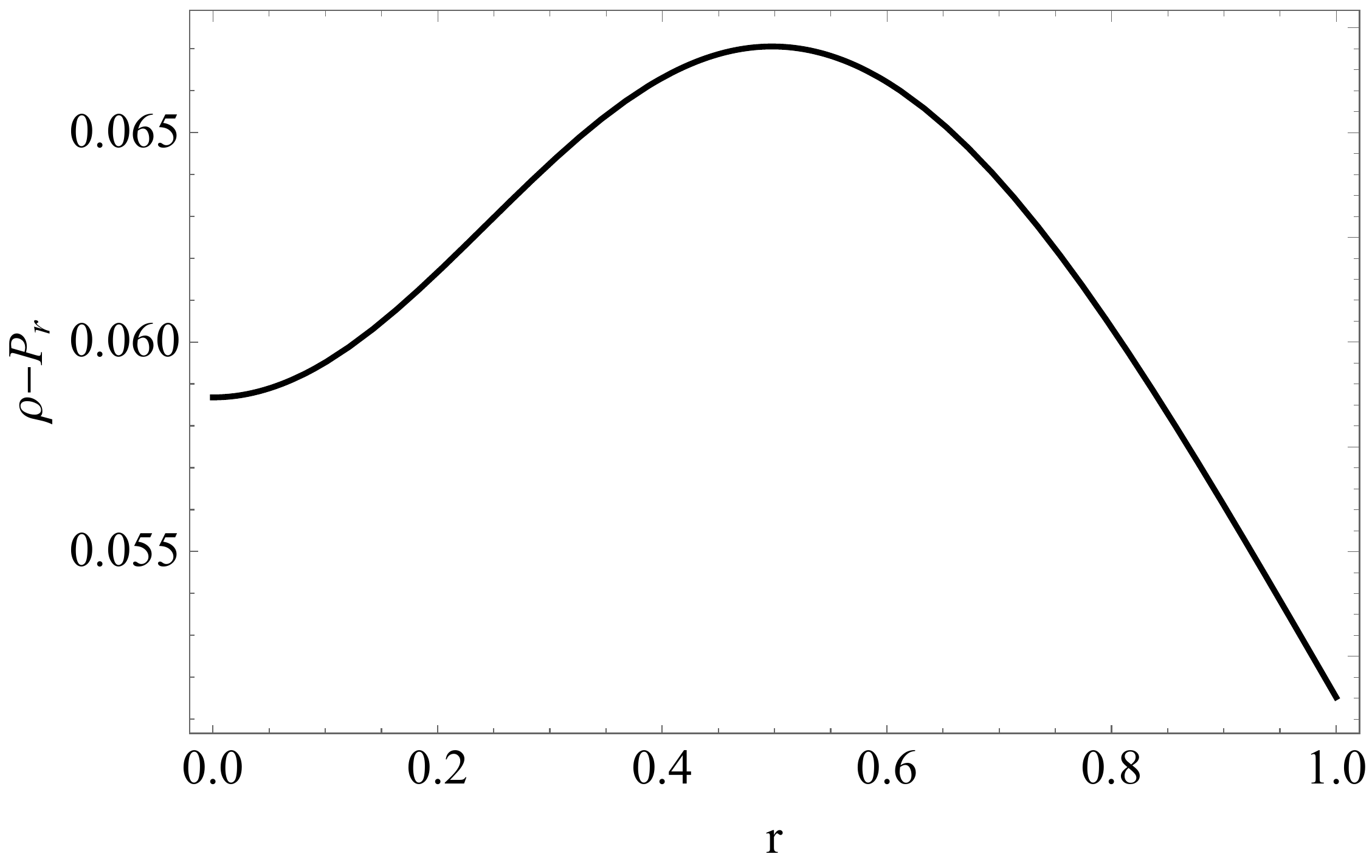}
\caption{\label{non_ch_dec1}
$\rho-P_{r}$ as a function of the radial coordinate $r$ for the un--charged Durgapal V (vanishing complexity) solution with $R=1$. 
}
\end{figure}

\begin{figure}[h!]
\centering
\includegraphics[scale=0.38]{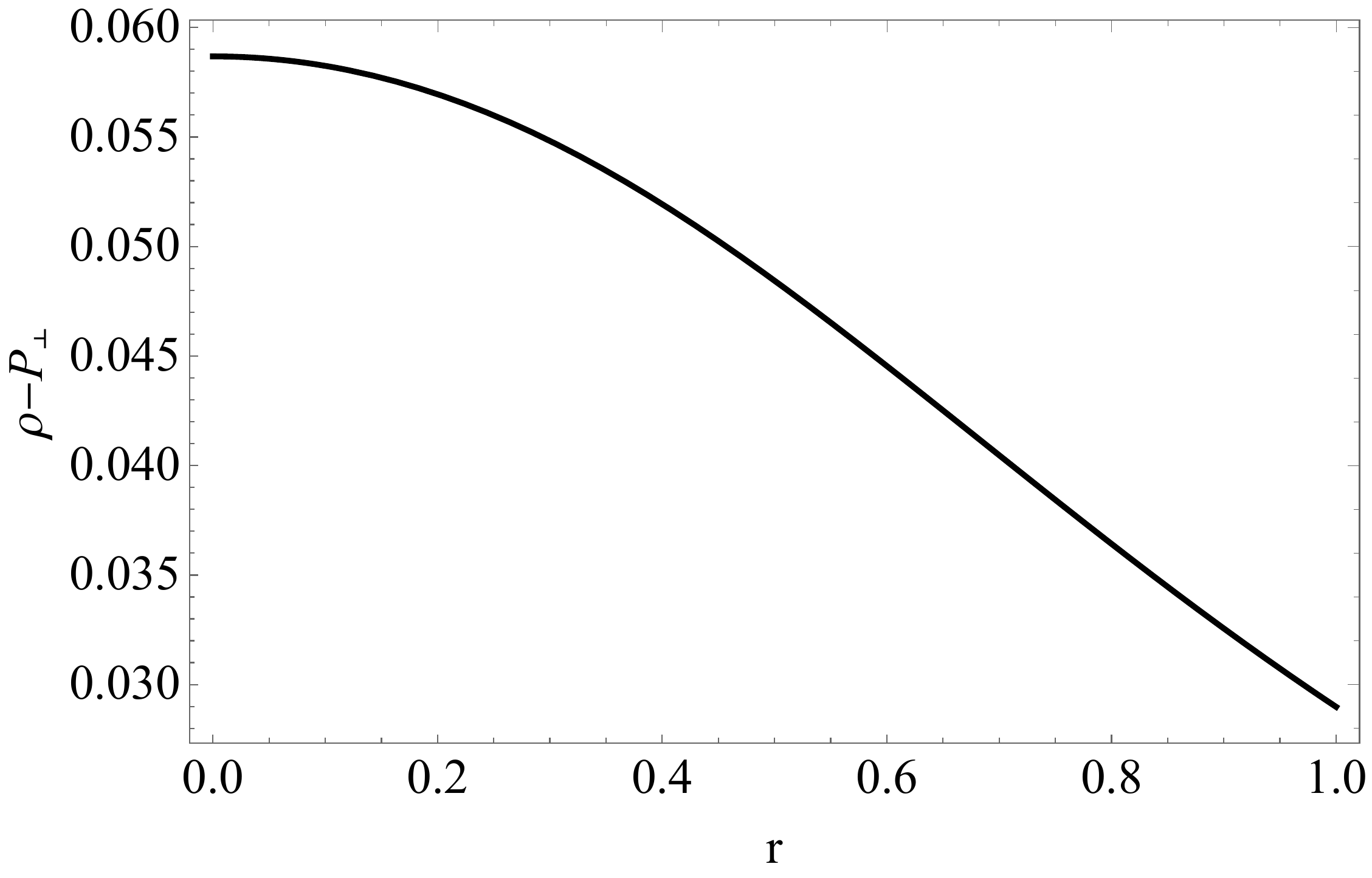}
\caption{\label{non_ch_dec2}
$\rho-P_{\perp}$ as a function of the radial coordinate $r$ ffor the un--charged Durgapal V (vanishing complexity) solution with $R=1$. 
}
\end{figure}

We continue showing the graphs corresponding to the matter sector, composed by its thermodynamic variables ($\rho$, $P_r$ and $P_{\perp}$) in figures \ref{densidad2}, \ref{pradial2} and \ref{ptangencial2} respectively, for the un--charged Durgapal V (vanishing complexity) solution, as well as the behavior of the resulting anisotropy $\Delta = P_{\perp} - P_r$ exposed in figure \ref{anisotropia2}. With this done, it is straightforward to see that the dominant energy condition is trivially satisfied (since $\rho$ $\geq$ $P_{\perp}$ and obviously $\rho$ $\geq$ $P_{r}$) as shown in the respective figures \ref{non_ch_dec1} and \ref{non_ch_dec2}.

\section{Charged Durgapal V solution}

Finally, to complete this work, we shall consider the charged Durgapal V solution in conjunction with the choice of a particular anisotropy function (proportional to $q^2$) and the condition for the vanishing of the complexity parameter $Y_{TF} = 0$. Again, the metric seed $g_{tt}$
component is
\begin{eqnarray}\label{dur_nu_v}
e^{\nu}=A \left(1+c r^2\right)^5\,,
\end{eqnarray}
and the vanishing complexity condition (\ref{YTF=0}) leads to
\begin{eqnarray}\label{dur_lambda_v}
e^{\lambda}=(1+c r^{2})^{3}.
\end{eqnarray}
Replacing (\ref{dur_nu_v}) and (\ref{dur_lambda_v}) in (\ref{e1}), (\ref{e2}) and (\ref{e3}) we obtain
\begin{eqnarray}
\rho &=& -\frac{q^2}{8 \pi  r^4} 
+ \frac{c \left(9 + 6cr^2 + 4c^2r^4 + c^3r^6\right)}{8 \pi  \left(1 + c r^2\right)^4}\;, \\
P_{r} &=& \frac{q^2}{8 \pi  r^4} + \frac{c \left( 7 - 6cr^2 - 4c^2r^4 - c^3r^6 \right)}{8 \pi  \left(1 + c r^2\right)^4}\;,\label{ch_Pr_V}\\
P_{\perp} &=& -\frac{q^2}{8 \pi  r^4}+\frac{7 c}{8 \pi  \left(1 + c r^2 \right)^4}.\label{ch_Pt_V}
\end{eqnarray}
From the previous equations, it is inferred that it is possible to absorb the electric charge in the thermodynamic variables of the matter sector by redefining effective variables, this constitutes a common procedure (see for example, \cite{Herrera:2011cr}). Just as in the previous treatment (charged Durgapal IV model) we use the anisotropy given by Eq. \eqref{ch_ani_IV}. Then, introducing the expressions (\ref{ch_Pr_V}) and (\ref{ch_Pt_V}) for both pressures in the left side of (\ref{ch_ani_IV}) we find the interior charge function, given by
\begin{eqnarray}
q=\frac{c r^3 \sqrt{6 + 4cr^2 + c^2r^4}}{\sqrt{2(1+\alpha)} \left(1 + c r^2\right)^{2}}.
\end{eqnarray}
This function, like in the Durgapal charged IV model, is inversely proportional to $\alpha$. This dependence of the internal electric field function with the parameter that modulates the local anisotropy has been reported before \cite{Gomez-Leyton:2020kfw}. 

Finally we apply the matching conditions \eqref{MC01}, \eqref{MC02}, \eqref{MC03} and \eqref{MC04} with the exterior Reissner--Nordstr\"om solution and we find that
\begin{eqnarray}
A &=& \frac{1}{(1 + cR^2)^8}\;, \\
M &=& \frac{M}{8}\Big\{ 4 - \frac{4}{(1+cR^2)^3} \nonumber \\
& & +
\frac{c^4R^{10}[6+cR^2(4+cR^2)]^2}{(1+\alpha)^2(1+cR^2)^8} \Big\}\;,\\
Q &=& \frac{c R^3 \sqrt{6 + 4cR^2 + c^2R^4}}{\sqrt{2(1+\alpha)} \left(1 + c R^2\right)^{2}}\;,\\
c &=& \frac{\Gamma}{3R^6(1+2\alpha)} - \frac{2R^2(1+2\alpha)}{3\Gamma}
- \frac{4}{3R^2}\; ,
\end{eqnarray}
with
\begin{eqnarray}
\Gamma &=& \Big[ R^{12}(1+2\alpha)^2 (233+277\alpha)  \nonumber\\ 
& & + 3 \sqrt{3R^{24}(1+2\alpha)^4[2011 + 
\alpha(4782+2843\alpha)]} \Big]^{1/3}.\nonumber\\
\end{eqnarray}
Note that in order to fix a specific solution, values must be given to $R$ and $\alpha$. If we turn off the interior charge function $q=0$, we obtain the un--charged anisotropic Durgapal V solution given in the previous case; but when there is indeed a presence of electric interior charge, the $\alpha$ parameter is used to control the anisotropy of the system, even to turn it off. The behavior for this anisotropic--charged Durgapal V model is in essence very similar, in practically all aspects, to that discussed for its Durgapal IV analog, so we proceed to summarize the most relevant issues.

\begin{figure}[h!]
\centering
\includegraphics[scale=0.5]{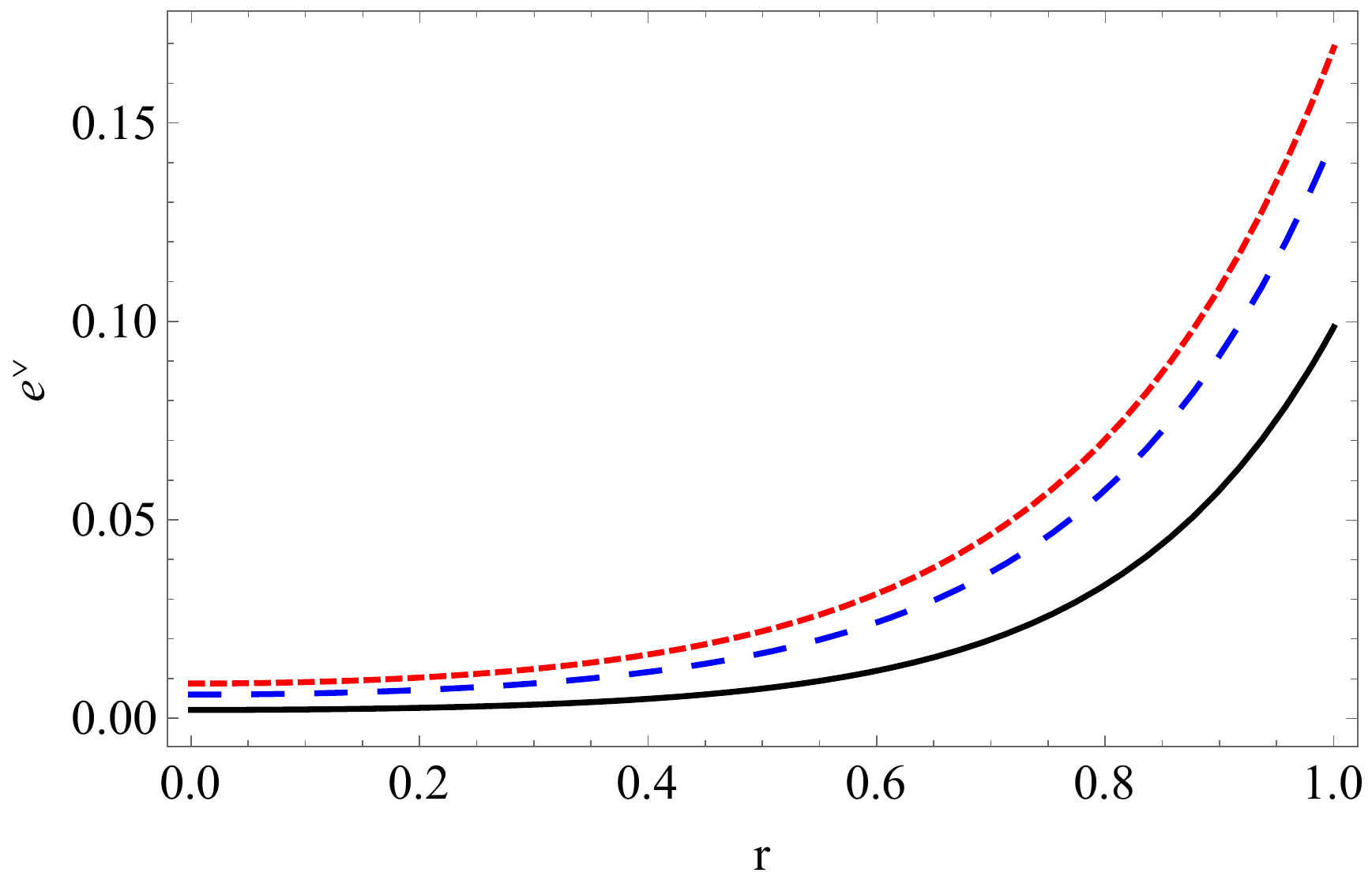}
\caption{\label{ch_enu_V}
Metric component $e^{\nu}$ as a function of the radial coordinate $r$ with $R=1$ and $\alpha=0$ (black line), 
$\alpha=1$ (blue dashed line) and $\alpha=3$ (red dotted line) for the charged Durgapal V anisotropic solution. 
}
\end{figure}

\begin{figure}[h!]
\centering
\includegraphics[scale=0.5]{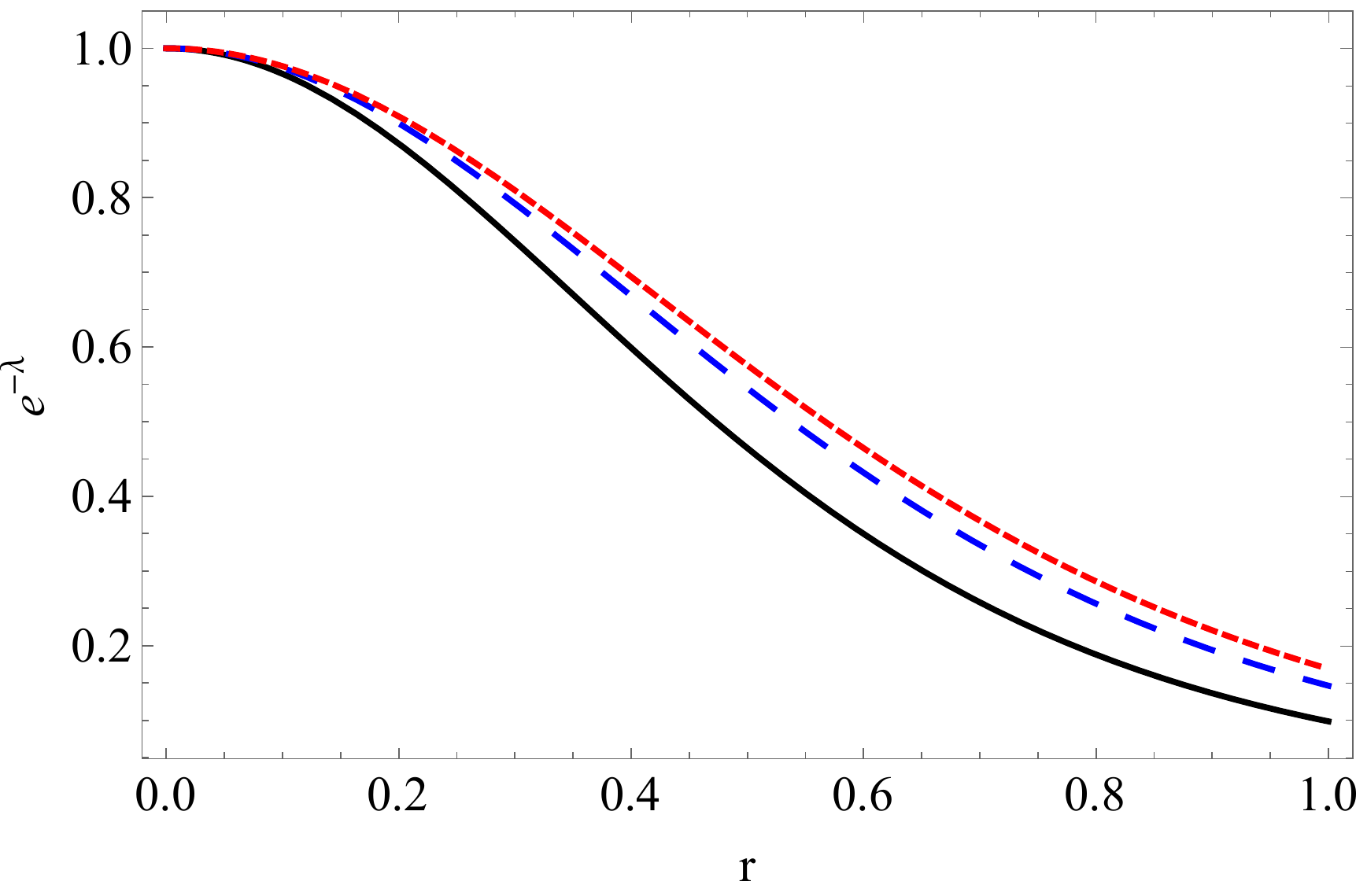}
\caption{\label{ch_elambda_V}
Metric component $e^{-\lambda}$ as a function of the radial coordinate $r$ with $R=1$ and $\alpha=0$ (black line), 
$\alpha=1$ (blue dashed line) and $\alpha=3$ (red dotted line) for the charged Durgapal V anisotropic solution. 
}
\end{figure}

In figures \ref{ch_enu_V} and \ref{ch_elambda_V} its exposed the behavior of metric function variables with respect to the radial coordinate $r$ for the anisotropic Durgapal V charged ($q\ne 0$) model, with $R=1$ and different values for the parameter $\alpha$ (that implies different anisotropies). Both functions are positive, finite and free of singularities, as they should be for a physical accepted solution. Both metric functions increase in value with increasing anisotropy as determined by the parameter $\alpha$. 

\begin{figure}[h!]
\centering
\includegraphics[scale=0.5]{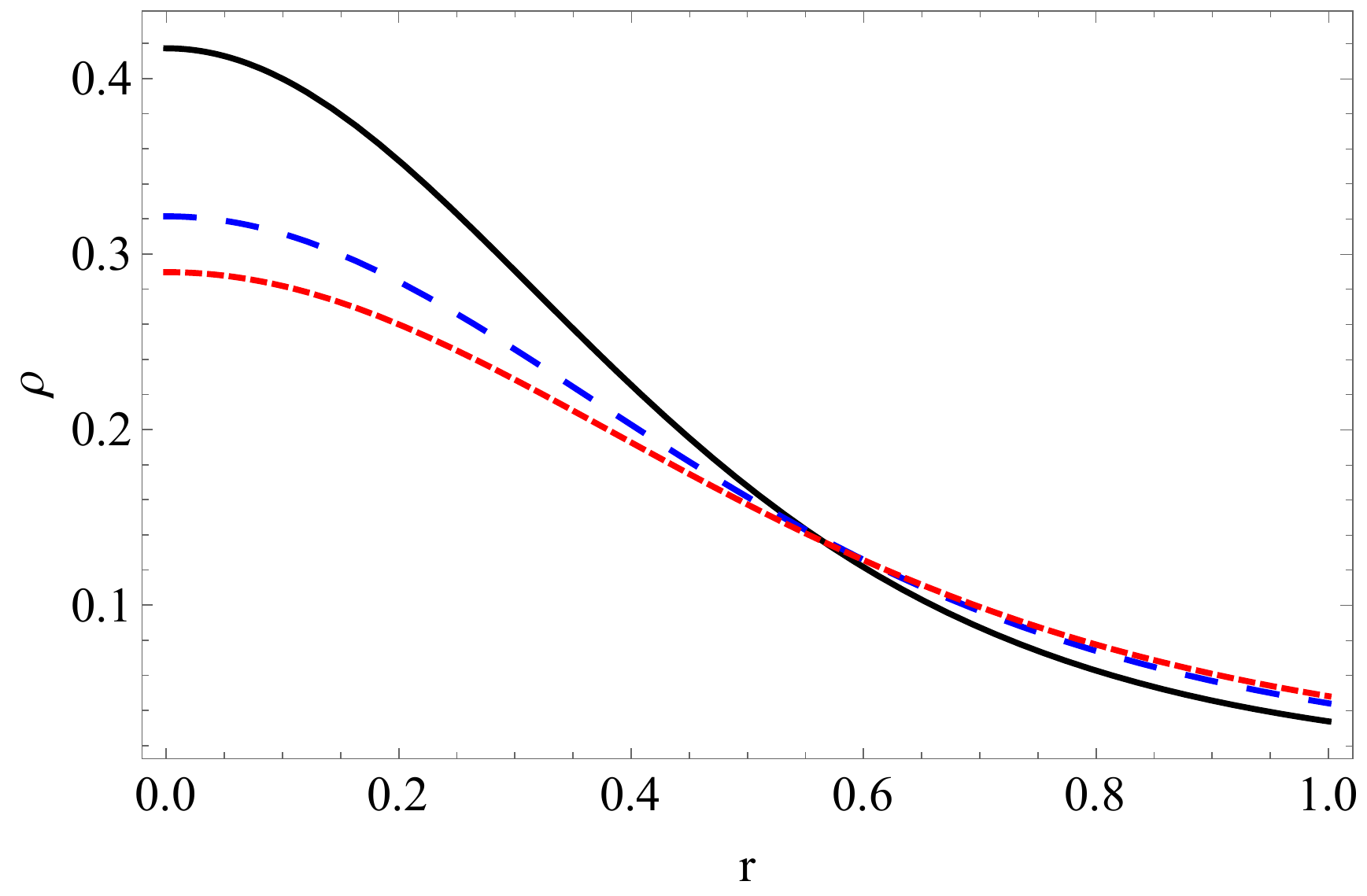}
\caption{\label{ch_densidad_V}
$\rho$ as a function of the radial coordinate $r$ with  $R=1$ and $\alpha=0$ (black line), 
$\alpha=1$ (blue dashed line) and $\alpha=3$ (red dotted line) for the charged Durgapal V anisotropic solution. 
}
\end{figure}

\begin{figure}[h!]
\centering
\includegraphics[scale=0.5]{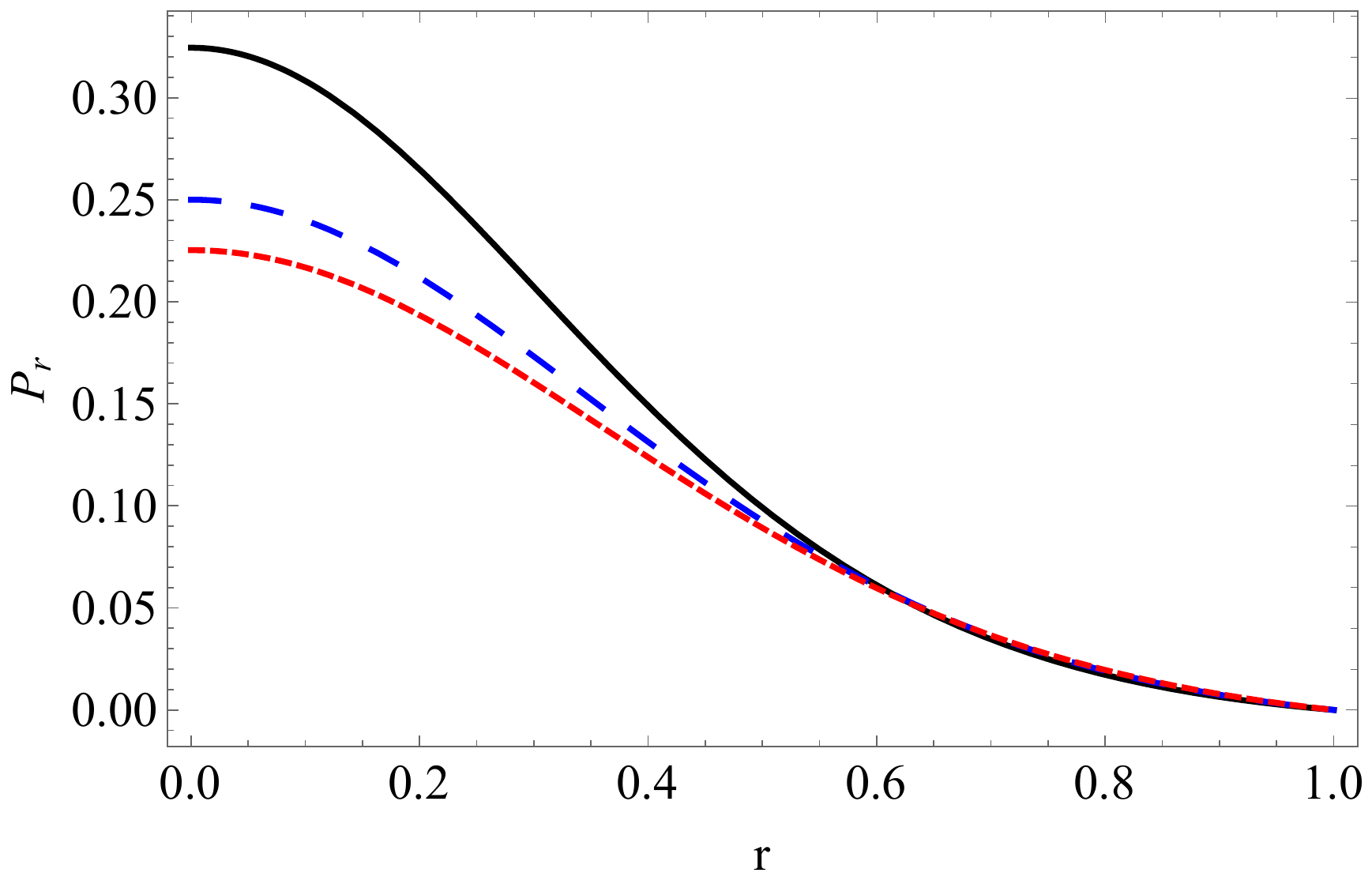}
\caption{\label{ch_pradial_V}
$P_{r}$ as a function of the radial coordinate $r$ with $R=1$ and $\alpha=0$ (black line), 
$\alpha=1$ (blue dashed line) and $\alpha=3$ (red dotted line) for the charged Durgapal V anisotropic solution. 
}
\end{figure}

\begin{figure}[h!]
\centering
\includegraphics[scale=0.5]{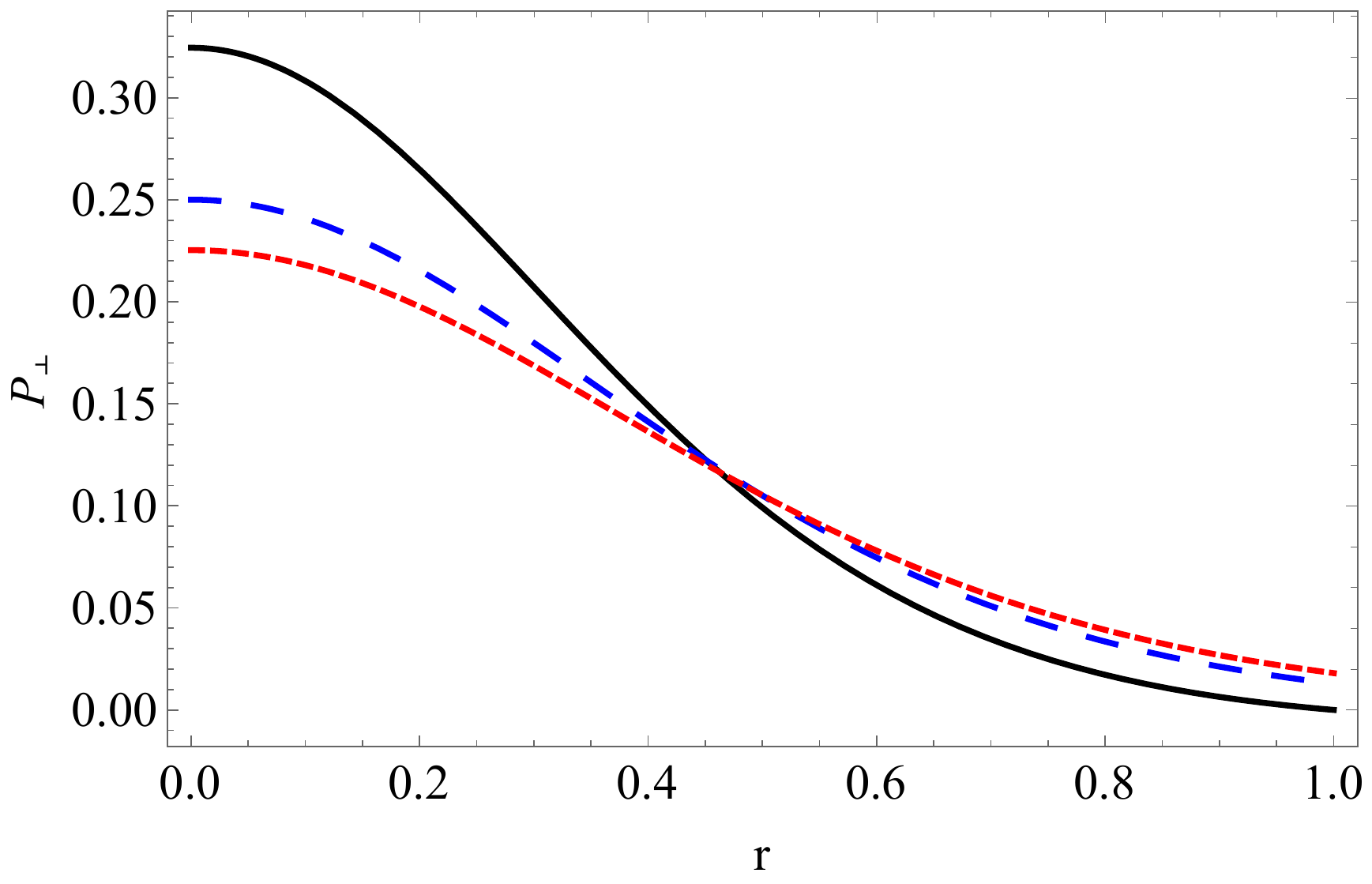}
\caption{\label{ch_ptangencial_V}
$P_{\perp}$ as a function of the radial coordinate $r$ with $R=1$ and $\alpha=0$ (black line), 
$\alpha=1$ (blue dashed line) and $\alpha=3$ (red dotted line) for the charged Durgapal V anisotropic solution. 
}
\end{figure}

\begin{figure}[h!]
\centering
\includegraphics[scale=0.5]{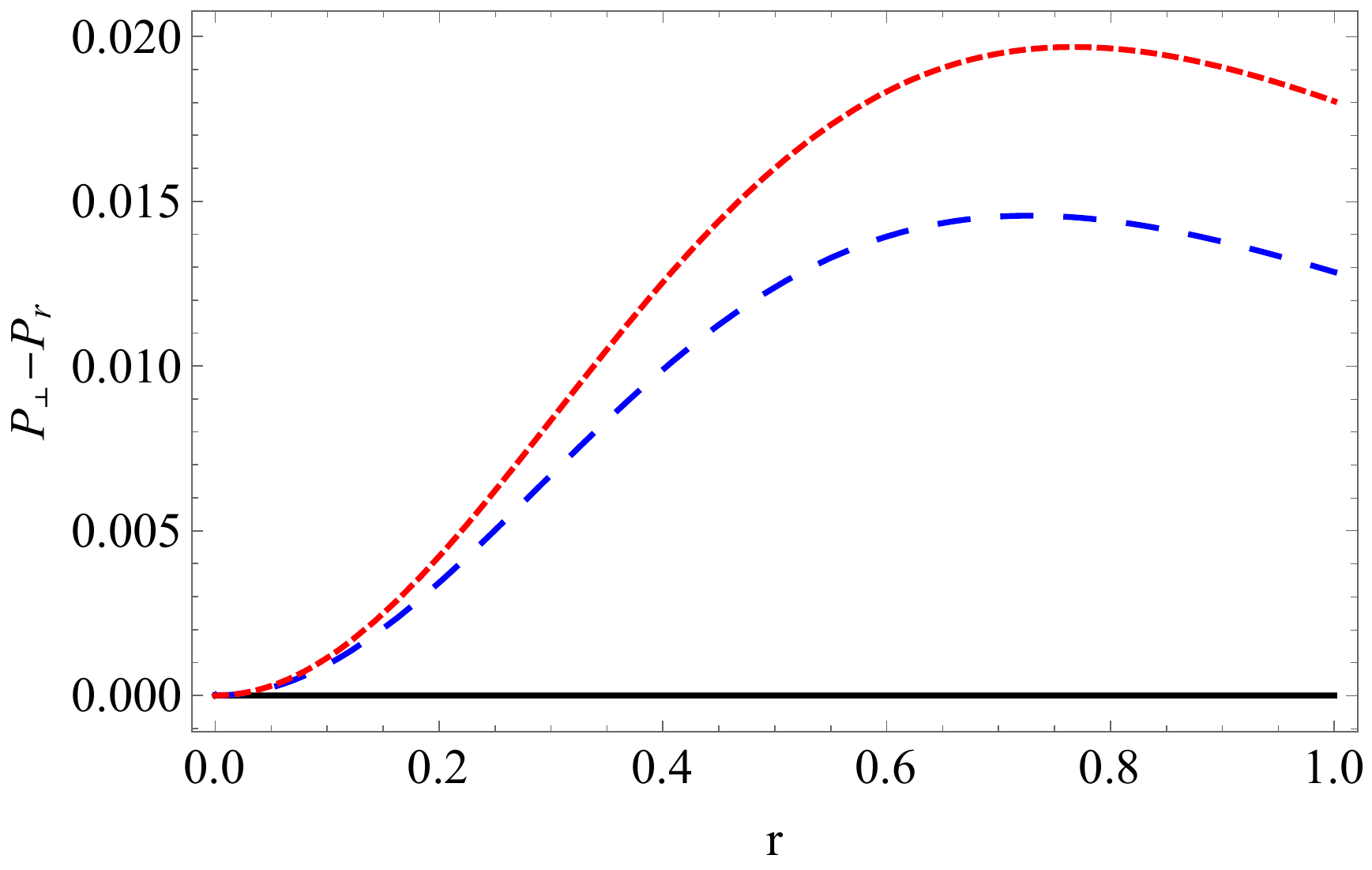}
\caption{\label{ch_anisotropia_V}
$P_{\perp}-P_{r}$ as a function of the radial coordinate $r$ with $R=1$ and $\alpha=0$ (black line), 
$\alpha=1$ (blue dashed line) and $\alpha=3$ (red dotted line) for the charged Durgapal V anisotropic solution. 
}
\end{figure}

\begin{figure}[h!]
\centering
\includegraphics[scale=0.5]{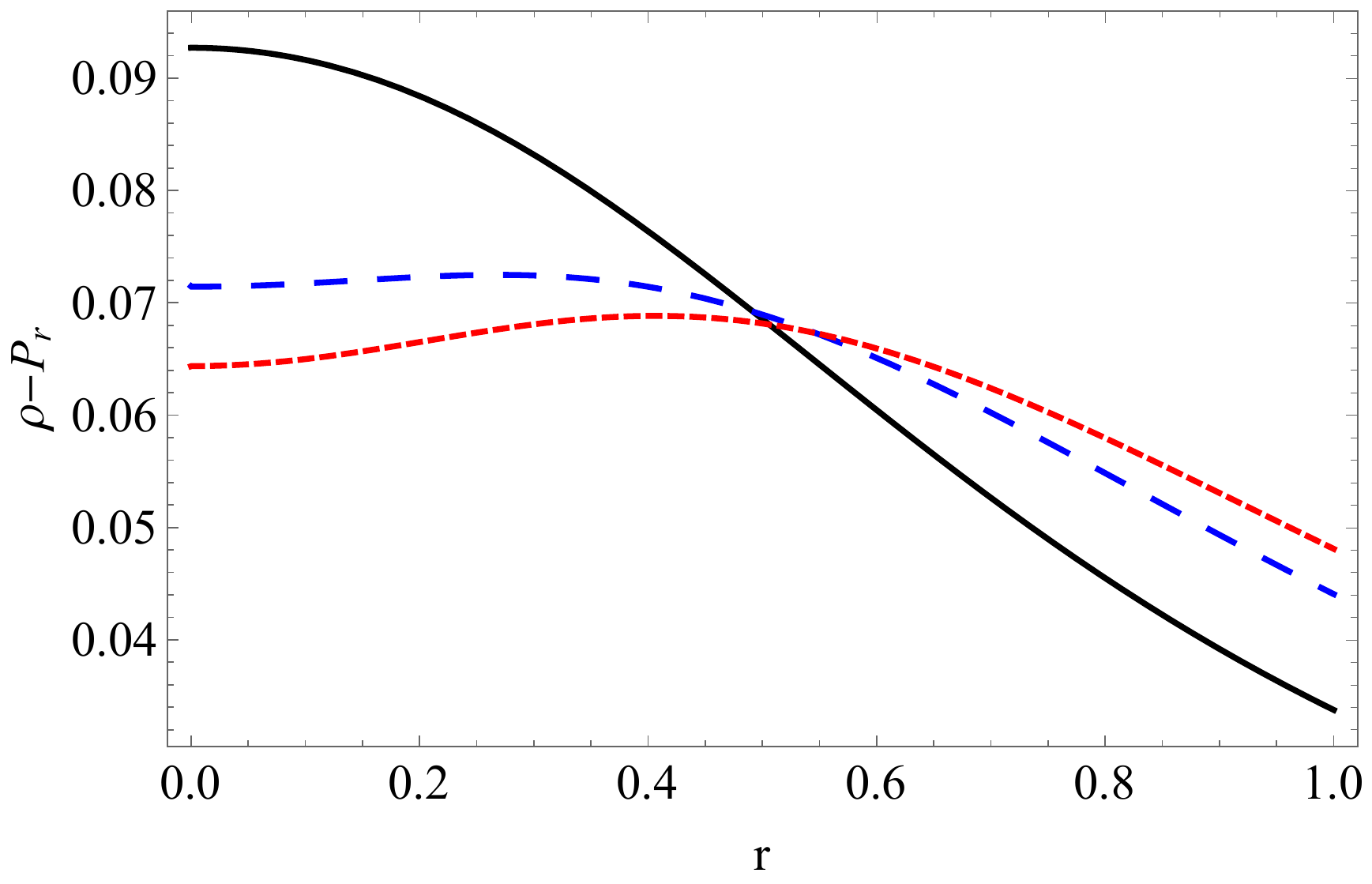}
\caption{\label{ch_dec1_V}
$\rho-P_{r}$ as a function of the radial coordinate $r$ with $R=1$ and $\alpha=0$ (black line), 
$\alpha=1$ (blue dashed line) and $\alpha=3$ (red dotted line) for the charged Durgapal V anisotropic solution.
}
\end{figure}

\begin{figure}[h!]
\centering
\includegraphics[scale=0.5]{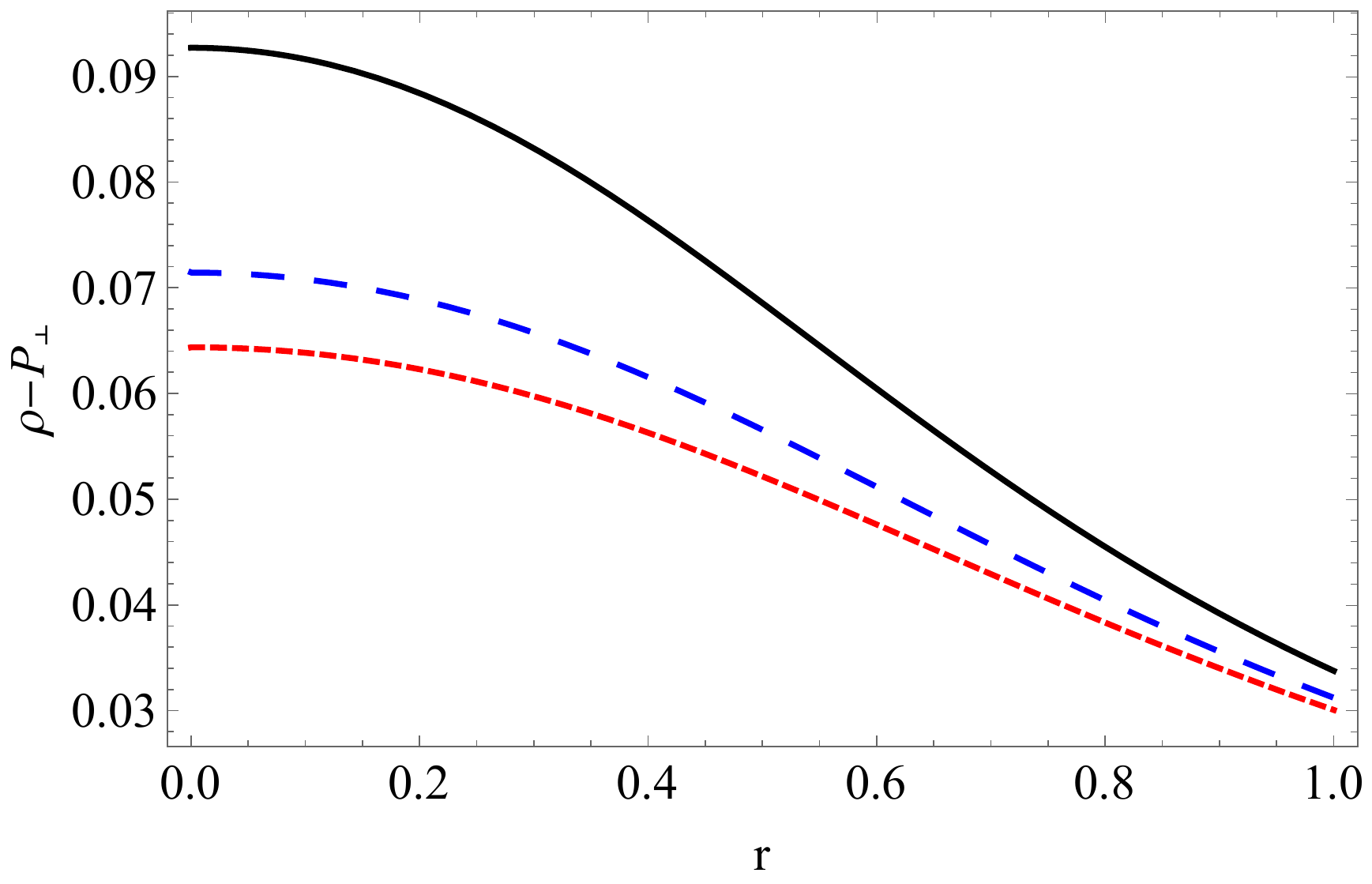}
\caption{\label{ch_dec2_V}
$\rho-P_{\perp}$ as a function of the radial coordinate $r$ with $R=1$ and $\alpha=0$ (black line), 
$\alpha=1$ (blue dashed line) and $\alpha=3$ (red dotted line) for the charged Durgapal V anisotropic solution. 
}
\end{figure}

In figures \ref{ch_densidad_V}, \ref{ch_pradial_V} and \ref{ch_ptangencial_V} we show the thermodynamic variables that constitute our stellar compact fluid, the density and pressures behave as expected so the inherited anisotropy present in the system (figure \ref{ch_anisotropia_V}) will do as well. In the deepest regions of the stellar body ($r < 0.5$) all quantities that make up the matter sector decrease in value with the increasing of the anisotropy via the $\alpha$ parameter. Then, when approaching the surface, the same figures show the opposite behavior (even all curves tend to converge). Consistent and similar results were found for the previous charged case (Durgapal IV). Furthermore, it is straightforward to observe that the energy conditions are correctly met as it is exposed in figures \ref{ch_dec1_V} and \ref{ch_dec2_V}.

\begin{figure}[h!]
\centering
\includegraphics[scale=0.5]{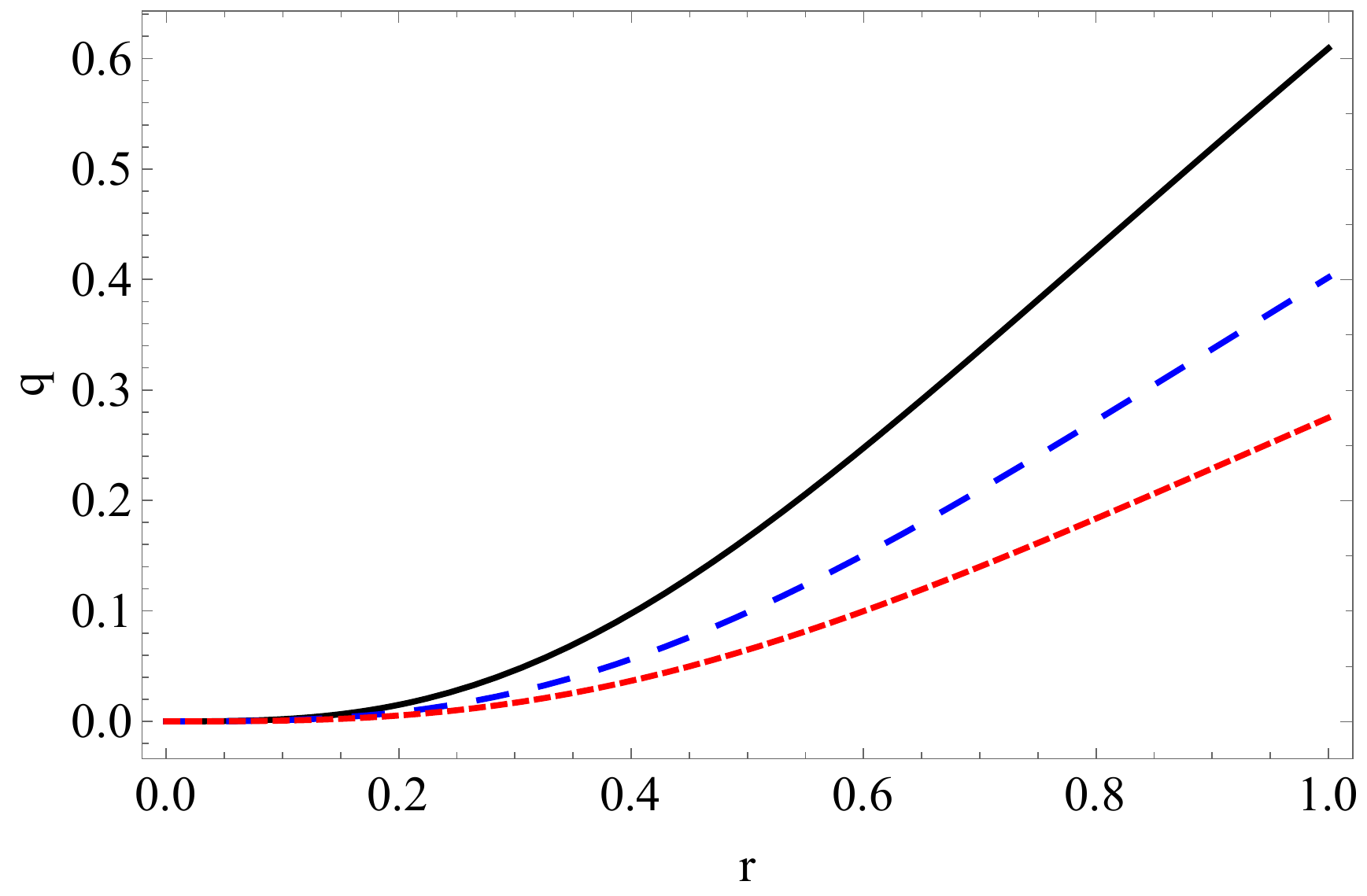}
\caption{\label{ch_carga_V}
Electric interior charge $q$ as a function of the radial coordinate $r$ with $R=1$ and $\alpha=0$ (black line), 
$\alpha=1$ (blue dashed line) and $\alpha=3$ (red dotted line) for the charged Durgapal V anisotropic solution. 
}
\end{figure}

Finally, the electric charge (or electric field) trend against the radial coordinate $r$ for the latter case, is shown in figure \ref{ch_carga_V}. As in the previous model containing electric charge, our results are totally consistent. Both electric charge $q(r)$ (and electric field $E(r)$) must be strictly positive and increasing functions with radius, meaning that at the origin both must be null (Gauss's law) i.e., $q(0) = E(0) = 0$. On the surface of the stellar object it is equivalent to the total charge $Q$. As before, the same decrease for the interior electric charge with increasing anisotropy is observed.

\section{Discussion}

Einstein's system of field equations in the case of a spherically symmetric static, anisotropic fluid form a system of three differential equations for five unknown quantities, the two metric functions $\nu $, $\lambda$ and the three thermodynamic variables (that constitute the so-called material sector): the energy density $\rho$, radial $P_r$ and tangential pressure $P_{\perp}$ of the fluid used to model the interior of the relativistic object. If we also consider an electrically charged fluid, we will therefore have an additional unknown variable (the charge or electric field within the fluid distribution). Accordingly, we shall need to provide extra information to be able to build a consistent model. One way to achieve this, allows us to use a well known solution as a {\it seed} to generate new ones. Unlike many previous works, where the radial component of the metric (mass function) is used as the seed solution \cite{herrera2018new}, we base the present work using the temporal component of the metric for Durgapal IV and V models, as a seed, and explore the possibility of finding solutions to Einstein's equations. If we consider the un-charged (charged) case we shall need still another (two more) condition (s) in order to solve the system. In this work, together with the Durgapal IV and V temporal metric as a seed solution we impose the vanishing of the complexity parameter $Y_{TF}=0$ (known as the vanishing complexity condition) that represents a non-local equation of state that allows to obtain a relationship between the metric functions. In general, setting some value for the complexity factor (for example, a system with vanishing complexity), works like an equation of state that leads to close the Einstein's field equations. Besides considering anisotropy as a quantity playing a fundamental role in relativistic compact stars, we explored the effect of this other interesting physical quantity that involves both the anisotropy and the gradients in the energy density (density contrast), namely the complexity factor.

The new concept of complexity for static spherically symmetric relativistic fluid distributions, arises from the basic assumption that the less complex systems corresponds to an homogeneous (in the energy density) fluid distribution with isotropic pressure \cite{herrera2018new}. In the case of fluid distributions the variable measuring the complexity of the fluid (the complexity factor $Y_{TF}$) appears in the trace-free part of the orthogonal splitting of the electric Riemann tensor (in vacuum the Riemann tensor and the Weyl tensor are the same) \cite{bel1961inductions, herrera2009structure, Gomez-Lobo:2007mbg}. If we consider a spherically symmetric static distribution of fluid, the magnetic part of the Weyl tensor is zero, so the process starts by calculating the scalar functions defining the electric part of the Weyl tensor. The scalar function $Y_{TF}$ contains contributions from the energy density inhomogeneity and the local pressure anisotropy, combined in a very specific way so it measures the departure from the value of the Tolman mass of a homogeneous and isotropic fluid and allows finding a great variety of models with vanishing complexity ($Y_{TF}=0$) different from the trivial case (perfect fluid). In the case that we have a electric charged fluid it is useful to introduce ``effective'' variables that are just the corresponding ordinary thermodynamic variables with the contributions of electric charge included (see \cite{Herrera:2011cr}). The remarkable fact emerging from this is that the charge contribution is always absorbed into the effective variables. In the absence of electrical charge, the structure scalars are obtained just replacing the effective variables by the corresponding ordinary ones \cite{Herrera:2011cr}, question that allows us to define complexity in the same way.

In this work, the equation of state resulting from canceling the structure scalar $Y_{TF}$ represents the other useful piece of information necessary for the consolidation of the models (in the case of configurations with electrical charge, it is also necessary to choose an anisotropy function). We analyzed the plausibility of each model based on the physical conditions established for the existence of anisotropic compact stars. All the considered cases behaved as expected in terms of metric potentials, matter sector, anisotropy function, internal charge of the system, energy conditions, redshift and stability (although in these latter conditions we did not present the graphs).

It is worth mentioning that the new definition of complexity for self-gravitating relativistic fluids, has been extended to consider the dynamic spherically symmetric situation (non--static case) \cite{Herrera:2018czt, Yousaf:2020dkj} and cases where the spherical symmetry is broken: the axially symmetric static case \cite{Herrera:2019cbx} and applications for some particular cases of cylindrically symmetric fluid distributions may be also found \cite{Sharif:2018efi,Zubair:2020poe}. In reference \cite{Herrera:2018czt} was treated complexity for dynamical spherically symmetric dissipative self-gravitating fluid distributions and in \cite{Herrera:2019xte} the complexity for the Bondi metric \cite{Bondi:1962px}. Besides the fact that the Bondi metric covers a vast numbers of spacetimes, Minkowski spacetime, the static Weyl metrics, nonradiative nonstatic metrics, it has, among other things, the virtue of providing a clear and precise criterion for the existence of gravitational radiation. Finally, electrically charged fluid configurations have also been included (see references \cite{Herrera:2018czt, Sharif:2018pgq} for details).

It is important to highlight that the role of charge distribution in the stability of such configurations can be clearly exhibited by equation (\ref{m3}). In particular it is worth stressing the fact that electric charge, unlike pressure, does not always produce a ``regeneration effect'' (does not always increase the Tolman mass). This fact together with the presence of the Coulomb term in the full set of the equations required for a description of physically meaningful models of collapsing charged spheres indicates the relevance of the electric charge in the process of collapse and therefore in the stability of compact spheres in hydrostatic equilibrium. Before concluding this work, it is worth mentioning that a detailed analysis of the solution by fixing the parameters to describe real stars is compulsory. However, as this is out of the scope of this paper we left this analysis to a future work.

\bibliographystyle{ieeetr}
\bibliography{biblio.bib}

\begin{thebibliography}{10}

\bibitem{Delgaty:1998uy}
M.~S.~R. Delgaty and K.~Lake, ``{Physical acceptability of isolated, static,
  spherically symmetric, perfect fluid solutions of Einstein's equations},''
  {\em Comput. Phys. Commun.}, vol.~115, pp.~395--415, 1998.

\bibitem{Herrera:1997plx}
L.~Herrera and N.~O. Santos, ``{Local anisotropy in self-gravitating
  systems},'' {\em Phys. Rept.}, vol.~286, pp.~53--130, 1997.

\bibitem{Herrera:2004xc}
L.~Herrera, A.~Di~Prisco, J.~Martin, J.~Ospino, N.~O. Santos, and O.~Troconis,
  ``{Spherically symmetric dissipative anisotropic fluids: A General study},''
  {\em Phys. Rev. D}, vol.~69, p.~084026, 2004.

\bibitem{Herrera:2007kz}
L.~Herrera, J.~Ospino, and A.~Di~Prisco, ``{All static spherically symmetric
  anisotropic solutions of Einstein's equations},'' {\em Phys. Rev. D},
  vol.~77, p.~027502, 2008.

\bibitem{Glass:2013nsa}
E.~N. Glass, ``{Generating Anisotropic Collapse and Expansion Solutions of
  Einstein's Equations},'' {\em Gen. Rel. Grav.}, vol.~45, pp.~2661--2670,
  2013.

\bibitem{Ovalle:2017wqi}
J.~Ovalle, R.~Casadio, R.~da~Rocha, and A.~Sotomayor, ``{Anisotropic solutions
  by gravitational decoupling},'' {\em Eur. Phys. J. C}, vol.~78, no.~2,
  p.~122, 2018.

\bibitem{Ovalle:2019lbs}
J.~Ovalle, C.~Posada, and Z.~Stuchl\'\i{}k, ``{Anisotropic ultracompact
  Schwarzschild star by gravitational decoupling},'' {\em Class. Quant. Grav.},
  vol.~36, no.~20, p.~205010, 2019.

\bibitem{Ovalle:2017fgl}
J.~Ovalle, ``{Decoupling gravitational sources in general relativity: from
  perfect to anisotropic fluids},'' {\em Phys. Rev. D}, vol.~95, no.~10,
  p.~104019, 2017.

\bibitem{Tello-Ortiz:2020svg}
F.~Tello-Ortiz, M.~Malaver, A.~Rinc\'on, and Y.~Gomez-Leyton, ``{Relativistic
  anisotropic fluid spheres satisfying a non-linear equation of state},'' {\em
  Eur. Phys. J. C}, vol.~80, no.~5, p.~371, 2020.

\bibitem{Azmat:2021qig}
H.~Azmat and M.~Zubair, ``{An anisotropic version of Tolman VII solution in
  $f(R, T)$ gravity via gravitational decoupling MGD approach},'' {\em Eur.
  Phys. J. Plus}, vol.~136, no.~1, p.~112, 2021.

\bibitem{Zubair:2020lna}
M.~Zubair and H.~Azmat, ``{Anisotropic Tolman V Solution by Minimal
  Gravitational Decoupling Approach},'' {\em Annals Phys.}, vol.~420,
  p.~168248, 2020.

\bibitem{Schmidt:1995eh}
G.~D. Schmidt and P.~S. Smith, ``{A Search for magnetic fields among DA white
  dwarfs},'' {\em Astrophys. J.}, vol.~448, p.~305, 1995.

\bibitem{Reimers:1995ia}
D.~Reimers, S.~Jordan, D.~Koester, N.~Bade, T.~Kohler, and L.~Wisotzki,
  ``{Discovery of four white dwarfs with strong magnetic fields by the Hamburg
  / ESO survey},'' {\em Astron. Astrophys.}, vol.~311, pp.~572--578, 1996.

\bibitem{Martinez:2003dz}
A.~P. Martinez, H.~P. Rojas, and H.~J. Mosquera~Cuesta, ``{Magnetic collapse of
  a neutron gas: Can magnetars indeed be formed?},'' {\em Eur. Phys. J. C},
  vol.~29, pp.~111--123, 2003.

\bibitem{PerezMartinez:2007kw}
A.~Perez~Martinez, H.~Perez~Rojas, and H.~Mosquera~Cuesta, ``{Anisotropic
  Pressures in Very Dense Magnetized Matter},'' {\em Int. J. Mod. Phys. D},
  vol.~17, pp.~2107--2123, 2008.

\bibitem{Ferrer:2010wz}
E.~J. Ferrer, V.~de~la Incera, J.~P. Keith, I.~Portillo, and P.~L. Springsteen,
  ``{Equation of State of a Dense and Magnetized Fermion System},'' {\em Phys.
  Rev. C}, vol.~82, p.~065802, 2010.

\bibitem{Andersson:2004aa}
N.~Andersson, G.~L. Comer, and K.~Glampedakis, ``{How viscous is a superfluid
  neutron star core?},'' {\em Nucl. Phys. A}, vol.~763, pp.~212--229, 2005.

\bibitem{Sad:2007afd}
B.~A. Sa'd, I.~A. Shovkovy, and D.~H. Rischke, ``{Bulk viscosity of strange
  quark matter: Urca versus non-leptonic processes},'' {\em Phys. Rev. D},
  vol.~75, p.~125004, 2007.

\bibitem{Alford:2007pj}
M.~G. Alford and A.~Schmitt, ``{Bulk viscosity in 2SC and CFL quark matter},''
  {\em AIP Conf. Proc.}, vol.~964, no.~1, pp.~256--263, 2007.

\bibitem{Drago:2003wg}
A.~Drago, A.~Lavagno, and G.~Pagliara, ``{Bulk viscosity in hybrid stars},''
  {\em Phys. Rev. D}, vol.~71, p.~103004, 2005.

\bibitem{Jones:2001ya}
P.~B. Jones, ``{Bulk viscosity of neutron star matter},'' {\em Phys. Rev. D},
  vol.~64, p.~084003, 2001.

\bibitem{vanDalen:2003uy}
E.~N.~E. van Dalen and A.~E.~L. Dieperink, ``{Bulk viscosity in neutron stars
  from hyperons},'' {\em Phys. Rev. C}, vol.~69, p.~025802, 2004.

\bibitem{Bayin:1982vw}
S.~S. Bayin, ``{Anisotropic Fluid Spheres in General Relativity},'' {\em Phys.
  Rev. D}, vol.~26, p.~1262, 1982.

\bibitem{DiPrisco:1997tw}
A.~Di~Prisco, L.~Herrera, and V.~Varela, ``{Cracking of Homogeneous
  Self-Gravitating Compact Objects Induced by Fluctuations of Local
  Anisotropy},'' {\em Gen. Rel. Grav.}, vol.~29, pp.~1239--1256, 1997.

\bibitem{Herrera:2020gdg}
L.~Herrera, ``{Stability of the isotropic pressure condition},'' {\em Phys.
  Rev. D}, vol.~101, no.~10, p.~104024, 2020.

\bibitem{Bonnor:1960a}
W.~B. {Bonnor}, ``{The mass of a static charged sphere},'' {\em Zeitschrift fur
  Physik}, vol.~160, pp.~59--65, Feb. 1960.

\bibitem{Florides_1983}
P.~S. Florides, ``The complete field of charged perfect fluid spheres and of
  other static spherically symmetric charged distributions,'' {\em Journal of
  Physics A: Mathematical and General}, vol.~16, pp.~1419--1433, may 1983.

\bibitem{Ray:2003gt}
S.~Ray, A.~L. Espindola, M.~Malheiro, J.~P.~S. Lemos, and V.~T. Zanchin,
  ``{Electrically charged compact stars and formation of charged black
  holes},'' {\em Phys. Rev. D}, vol.~68, p.~084004, 2003.

\bibitem{Ghezzi:2005iy}
C.~R. Ghezzi, ``{Relativistic structure, stability and gravitational collapse
  of charged neutron stars},'' {\em Phys. Rev. D}, vol.~72, p.~104017, 2005.

\bibitem{Boehmer:2007gq}
C.~G. Boehmer and T.~Harko, ``{Minimum mass-radius ratio for charged
  gravitational objects},'' {\em Gen. Rel. Grav.}, vol.~39, pp.~757--775, 2007.

\bibitem{Giuliani:2007zza}
A.~Giuliani and T.~Rothman, ``{Absolute stability limit for relativistic
  charged spheres},'' {\em Gen. Rel. Grav.}, vol.~40, pp.~1427--1447, 2008.

\bibitem{Andreasson:2008xw}
H.~Andreasson, ``{Sharp bounds on the critical stability radius for
  relativistic charged spheres},'' {\em Commun. Math. Phys.}, vol.~288, p.~715,
  2009.

\bibitem{MafaTakisa:2019nkj}
P.~Mafa~Takisa, S.~D. Maharaj, and L.~L. Leeuw, ``{Effect of electric charge on
  conformal compact stars},'' {\em Eur. Phys. J. C}, vol.~79, no.~1, p.~8,
  2019.

\bibitem{Anninos:2001yb}
P.~Anninos and T.~Rothman, ``{Instability of extremal relativistic charged
  spheres},'' {\em Phys. Rev. D}, vol.~65, p.~024003, 2002.

\bibitem{Ivanov:2002jy}
B.~V. Ivanov, ``{Static charged perfect fluid spheres in general relativity},''
  {\em Phys. Rev. D}, vol.~65, p.~104001, 2002.

\bibitem{Barreto:2006cr}
W.~Barreto, B.~Rodriguez, L.~Rosales, and O.~Serrano, ``{Self-similar and
  charged radiating spheres: An Anisotropic approach},'' {\em Gen. Rel. Grav.},
  vol.~39, pp.~23--39, 2007.
\newblock [Erratum: Gen.Rel.Grav. 39, 537--538 (2007)].

\bibitem{Herrera:2018czt}
L.~Herrera, A.~Di~Prisco, and J.~Ospino, ``{Definition of complexity for
  dynamical spherically symmetric dissipative self-gravitating fluid
  distributions},'' {\em Phys. Rev. D}, vol.~98, no.~10, p.~104059, 2018.

\bibitem{Sharif:2018pgq}
M.~Sharif and I.~I. Butt, ``{Complexity Factor for Charged Spherical System},''
  {\em Eur. Phys. J. C}, vol.~78, no.~8, p.~688, 2018.

\bibitem{Zubair:2021zqs}
M.~Zubair, M.~Amin, and H.~Azmat, ``{Anisotropic charged Heintzmann solution
  using gravitational decoupling through extended geometric deformation
  approach},'' {\em Phys. Scripta}, vol.~96, no.~12, 2021.

\bibitem{Azmat:2021kmv}
H.~Azmat and M.~Zubair, ``{Anisotropic counterpart of charged Durgapal V
  perfect fluid sphere},'' {\em Int. J. Mod. Phys. D}, vol.~30, no.~15,
  p.~2150115, 2021.

\bibitem{Herrera:2001vg}
L.~Herrera, A.~Di~Prisco, J.~Ospino, and E.~Fuenmayor, ``{Conformally flat
  anisotropic spheres in general relativity},'' {\em J. Math. Phys.}, vol.~42,
  pp.~2129--2143, 2001.

\bibitem{herrera2014conformally}
L.~Herrera, A.~Di~Prisco, W.~Barreto, and J.~Ospino, ``Conformally flat
  polytropes for anisotropic matter,'' {\em General Relativity and
  Gravitation}, vol.~46, no.~12, p.~1827, 2014.

\bibitem{karmarkar1948gravitational}
K.~Karmarkar, ``Gravitational metrics of spherical symmetry and class one,'' in
  {\em Proceedings of the Indian Academy of Sciences-Section A}, vol.~27,
  p.~56, Springer, 1948.

\bibitem{ramos2021class}
A.~Ramos, C.~Arias, E.~Fuenmayor, and E.~Contreras, ``Class i polytropes for
  anisotropic matter,'' {\em The European Physical Journal C}, vol.~81, no.~3,
  pp.~1--10, 2021.

\bibitem{tello2019anisotropic}
F.~Tello-Ortiz, S.~Maurya, A.~Errehymy, K.~N. Singh, and M.~Daoud,
  ``Anisotropic relativistic fluid spheres: an embedding class i approach,''
  {\em The European Physical Journal C}, vol.~79, no.~11, pp.~1--14, 2019.

\bibitem{ospino2020karmarkar}
J.~Ospino and L.~N{\'u}{\~n}ez, ``Karmarkar scalar condition,'' {\em The
  European Physical Journal C}, vol.~80, no.~2, pp.~1--9, 2020.

\bibitem{maurya2017anisotropic}
S.~Maurya and S.~Maharaj, ``Anisotropic fluid spheres of embedding class one
  using karmarkar condition,'' {\em The European Physical Journal C}, vol.~77,
  no.~5, pp.~1--13, 2017.

\bibitem{pant2021new}
N.~Pant, M.~Govender, and S.~Gedela, ``A new class of viable and exact
  solutions of efes with karmarkar conditions: an application to cold star
  modeling,'' {\em Research in Astronomy and Astrophysics}, vol.~21, no.~5,
  p.~109, 2021.

\bibitem{baskey2021analytical}
L.~Baskey, S.~Das, and F.~Rahaman, ``An analytical anisotropic compact stellar
  model of embedding class i,'' {\em Modern Physics Letters A}, vol.~36,
  no.~05, p.~2150028, 2021.

\bibitem{comp3}
S.~Lloyd and H.~Pagels, ``Complexity as thermodynamic depth,'' {\em Annals of
  physics}, vol.~188, no.~1, pp.~186--213, 1988.

\bibitem{com4}
J.~P. Crutchfield and K.~Young, ``Inferring statistical complexity,'' {\em
  Physical review letters}, vol.~63, no.~2, p.~105, 1989.

\bibitem{comp5}
P.~W. Anderson, ``Is complexity physics? is it science? what is it?,'' {\em
  Physics Today}, vol.~44, no.~7, p.~9, 1991.

\bibitem{comp7}
R.~Lopez-Ruiz, H.~L. Mancini, and X.~Calbet, ``A statistical measure of
  complexity,'' {\em Physics letters A}, vol.~209, no.~5-6, pp.~321--326, 1995.

\bibitem{comp8}
D.~P. Feldman and J.~P. Crutchfield, ``Measures of statistical complexity:
  Why?,'' {\em Physics Letters A}, vol.~238, no.~4-5, pp.~244--252, 1998.

\bibitem{comp10}
X.~Calbet and R.~L{\'o}pez-Ruiz, ``Tendency towards maximum complexity in a
  nonequilibrium isolated system,'' {\em Physical Review E}, vol.~63, no.~6,
  p.~066116, 2001.

\bibitem{comp11}
R.~G. Catal{\'a}n, J.~Garay, and R.~L{\'o}pez-Ruiz, ``Features of the extension
  of a statistical measure of complexity to continuous systems,'' {\em Physical
  Review E}, vol.~66, no.~1, p.~011102, 2002.

\bibitem{comp12}
J.~Sa{\~n}udo and R.~L{\'o}pez-Ruiz, ``Statistical complexity and
  fisher--shannon information in the h-atom,'' {\em Physics Letters A},
  vol.~372, no.~32, pp.~5283--5286, 2008.

\bibitem{comp13}
C.~Panos, N.~Nikolaidis, K.~C. Chatzisavvas, and C.~Tsouros, ``A simple method
  for the evaluation of the information content and complexity in atoms. a
  proposal for scalability,'' {\em Physics Letters A}, vol.~373, no.~27-28,
  pp.~2343--2350, 2009.

\bibitem{grunwald2003kolmogorov}
P.~D. Gr{\"u}nwald and P.~M. Vit{\'a}nyi, ``Kolmogorov complexity and
  information theory. with an interpretation in terms of questions and
  answers,'' {\em Journal of Logic, Language and Information}, vol.~12, no.~4,
  pp.~497--529, 2003.

\bibitem{chapman2017complexity}
S.~Chapman, H.~Marrochio, and R.~C. Myers, ``Complexity of formation in
  holography,'' {\em Journal of High Energy Physics}, vol.~2017, no.~1,
  pp.~1--61, 2017.

\bibitem{yang2020time}
R.-Q. Yang and K.-Y. Kim, ``Time evolution of the complexity in chaotic
  systems: a concrete example,'' {\em Journal of High Energy Physics},
  vol.~2020, no.~1906.02052, pp.~1--33, 2020.

\bibitem{herrera2018new}
L.~Herrera, ``New definition of complexity for self-gravitating fluid
  distributions: The spherically symmetric, static case,'' {\em Physical Review
  D}, vol.~97, no.~4, p.~044010, 2018.

\bibitem{bel1961inductions}
L.~Bel, ``Inductions {\'e}lectromagn{\'e}tique et gravitationnelle,'' in {\em
  Annales de l'institut Henri Poincar{\'e}}, vol.~17, pp.~37--57, 1961.

\bibitem{herrera2009structure}
L.~Herrera, J.~Ospino, A.~Di~Prisco, E.~Fuenmayor, and O.~Troconis,
  ``{Structure and evolution of self-gravitating objects and the orthogonal
  splitting of the Riemann tensor},'' {\em Phys. Rev. D}, vol.~79, p.~064025,
  2009.

\bibitem{Gomez-Lobo:2007mbg}
A.~G.-P. Gomez-Lobo, ``{Dynamical laws of superenergy in General Relativity},''
  {\em Class. Quant. Grav.}, vol.~25, p.~015006, 2008.

\bibitem{schwarzschild1916gravitationsfeld}
K.~Schwarzschild, ``{\"U}ber das gravitationsfeld eines massenpunktes nach der
  einsteinschen theorie,'' {\em Sitzungsberichte der K{\"o}niglich
  Preu{\ss}ischen Akademie der Wissenschaften (Berlin}, pp.~189--196, 1916.

\bibitem{Herrera:1997si}
L.~Herrera, A.~Di~Prisco, J.~L. Hernandez-Pastora, and N.~O. Santos, ``{On the
  role of density inhomogeneity and local anisotropy in the fate of spherical
  collapse},'' {\em Phys. Lett. A}, vol.~237, pp.~113--118, 1998.

\bibitem{Hernandez_2004}
H.~Hernández and L.~A. Núñez, ``Nonlocal equation of state in anisotropic
  static fluid spheres in general relativity,'' {\em Canadian Journal of
  Physics}, vol.~82, p.~29–51, Jan 2004.

\bibitem{Bekenstein:1971ej}
J.~D. Bekenstein, ``{Hydrostatic Equilibrium and Gravitational Collapse of
  Relativistic Charged Fluid Balls},'' {\em Phys. Rev. D}, vol.~4,
  pp.~2185--2190, 1971.

\bibitem{Herrera:2011cr}
L.~Herrera, A.~Di~Prisco, and J.~Ibanez, ``{On the Role of Electric Charge and
  Cosmological Constant in Structure Scalars},'' {\em Phys. Rev. D}, vol.~84,
  p.~107501, 2011.

\bibitem{Israel:1966rt}
W.~Israel, ``{Singular hypersurfaces and thin shells in general relativity},''
  {\em Nuovo Cim. B}, vol.~44S10, p.~1, 1966.
\newblock [Erratum: Nuovo Cim.B 48, 463 (1967)].

\bibitem{Hernandez:2020pcn}
H.~Hern\'andez, D.~Su\'arez-Urango, and L.~A. N\'u\~nez, ``{Acceptability
  Conditions and Relativistic Barotropic Equations of State},'' {\em Eur. Phys.
  J. C}, vol.~81, no.~3, p.~241, 2021.

\bibitem{Suarez-Urango:2021cjy}
D.~Su\'arez-Urango, L.~A. N\'u\~nez, and H.~Hern\'andez, ``{Relativistic
  Anisotropic Polytropic Spheres: Physical Acceptability},'' 1 2021.

\bibitem{Suarez-Urango:2021mjg}
D.~Su\'arez-Urango, J.~Ospino, H.~Hern\'andez, and L.~A. N\'u\~nez,
  ``{Acceptability Conditions and Relativistic Anisotropic Generalized
  Polytropes},'' 4 2021.

\bibitem{Dey:2020fxm}
S.~Dey and B.~C. Paul, ``{Higher dimensional charged compact objects in
  Finch\textendash{}Skea geometry},'' {\em Class. Quant. Grav.}, vol.~37,
  no.~7, p.~075017, 2020.

\bibitem{Gomez-Leyton:2020kfw}
Y.~Gomez-Leyton, H.~Javaid, L.~S. Rocha, and F.~Tello-Ortiz, ``{Charged
  anisotropic compact objects obeying Karmarkar condition},'' {\em Phys.
  Scripta}, vol.~96, no.~2, p.~025001, 2021.

\bibitem{Yousaf:2020dkj}
Z.~Yousaf, M.~Z. Bhatti, and T.~Naseer, ``{Measure of complexity for dynamical
  self-gravitating structures},'' {\em Int. J. Mod. Phys. D}, vol.~29, no.~09,
  p.~2050061, 2020.

\bibitem{Herrera:2019cbx}
L.~Herrera, A.~Di~Prisco, and J.~Ospino, ``{Complexity factors for axially
  symmetric static sources},'' {\em Phys. Rev. D}, vol.~99, no.~4, p.~044049,
  2019.

\bibitem{Sharif:2018efi}
M.~Sharif and I.~I. Butt, ``{Complexity factor for static cylindrical
  system},'' {\em Eur. Phys. J. C}, vol.~78, no.~10, p.~850, 2018.

\bibitem{Zubair:2020poe}
M.~Zubair and H.~Azmat, ``{Complexity analysis of cylindrically symmetric
  self-gravitating dynamical system in f(R,T) theory of gravity},'' {\em Phys.
  Dark Univ.}, vol.~28, p.~100531, 2020.

\bibitem{Herrera:2019xte}
L.~Herrera, A.~Di~Prisco, and J.~Carot, ``{Complexity of the Bondi metric},''
  {\em Phys. Rev. D}, vol.~99, no.~12, p.~124028, 2019.

\bibitem{Bondi:1962px}
H.~Bondi, M.~G.~J. van~der Burg, and A.~W.~K. Metzner, ``{Gravitational waves
  in general relativity. 7. Waves from axisymmetric isolated systems},'' {\em
  Proc. Roy. Soc. Lond. A}, vol.~269, pp.~21--52, 1962.

\end{thebibliography}

\end{document}